\documentclass[aip,pop,12pt,reprint]{revtex4-1}

\usepackage{graphicx,bm,amsmath}

\let\marginparo\marginpar

\def\marginpar#1{\marginparo{\small#1}}

\newcommand*\degrees{}

\catcode`\@=11 

\newif\ifjournal

\newcommand*\Unskip{\unskip~}	

\newcommand*\Appabbrev{Appendix} 

\newcommand*\App[1]{\Appabbrev~\seco{#1}}

\newcommand*\eq[1]{\label{#1}}	
\newcommand*\Eq[1]{Eq.~(\ref{#1})}
\newcommand*\Eqs[1]{Eqs.~(\ref{#1})}
\newcommand*\EQo[1]{(\ref{#1})} 
\newcommand*\EQ[1]{\Unskip\EQo{#1}}
\newcommand*\Equation[1]{Equation~(\ref{#1})}	
\newcommand*\Equations[1]{Equations~(\ref{#1})}

\newcommand*\Fig[2][]{Fig.~\seco[#1]{Fg.#2}}

\newcommand*\Ref[1]{Ref.~\Onlinecite{#1}}

\newcommand*\REF[1]{\Unskip\Onlinecite{#1}}
\newcommand*\REFo[1]{{\let\ \relax\Onlinecite{#1}}} 

\newcommand*\Onlinecite{\onlinecite} 

\newcommand*\seco[2][]{\ref{#2}#1}	

\newcommand*\Sec[1]{Sec.~\seco{#1}}

\newcommand*\Ifarrayflag{\ifarrayflag}
\newcommand*\Ifletterflag{\ifletterflag}
\newcommand*\Ifbeginningeq{\ifbeginningeq}

\newcommand*\BE{\arrayflagfalse\beginningeqfalse\begin{equation}} 

\newcommand*\BEA{\arrayflagtrue\beginningeqfalse\begin{eqnarray}} 
\newcommand*\BA{\BEA} 
\def\BAams#1\EAams{\begin{align}#1\end{align}} 

\newcommand*\EEA{\end{eqnarray}}

\newcommand*\BAL[1][]{\letterflagtrue\st@rtarray[#1]}
\newcommand*\EAL{\end{eqnarray}\EM}

\newcommand*\EE{%
\Ifbeginningeq
	\beginningeqfalse
	\BE
\else
	\Endanequation
	\beginningeqtrue
\fi
\Ifletterflag
	\EM
\fi
}

\newcommand*\Endanequation{%
\Ifarrayflag
	\end{eqnarray}%
\else
	\end{equation}%
\fi
}

\newcommand*\BM[1][]{\begin{subequations}%
	\gdef\theletters{a}%
	\def\NP##1{##1%
		\ifx\df@label\@empty\else\@xp\ltx@label\@xp{\df@label}\fi
		\let\df@label\@empty
		\def\theequation{\theparentequation\theletters}%
		\stepcounter{equation}%
		\protected@edef\@currentlabel{\theparentequation\alph{equation}}%
		\xdef\theletters{\theletters,\alph{equation}}%
		}%
	\def\NQ##1{\xdef\theletters{\alph{equation}}%
		\NP{##1}}%
	\l@belletters{#1}%
	}

\newcommand*\l@belletters[1]{\def\Jtemp{#1}\ifx\Jtemp\empty\else\eq{#1}\fi}

\newcommand*\EM{\end{subequations}\COMMENT
	\letterflagfalse
	}

\newcommand*\NN{\nonumber}

\newcommand*\BI{\begin{itemize}}
\newcommand*\EI{\end{itemize}}

\newcommand*\verticalbar{|}

\newcommand*\e{\epsilon}
\newcommand*\eps{\varepsilon}
\newcommand*\g{\gamma}
\newcommand*\G{\Gamma}

\renewcommand*\k{\kappa}

\newcommand*\p{\phi}

\newcommand*\ph{\varphi}

\newcommand*\U{\Upsilon}
\newcommand*\w{\omega}

\newcommand*\z{\zeta}

\newcommand*\shortGreek{
	\renewcommand*\a{\alpha}
	\renewcommand*\b{\beta}
	\renewcommand*\c{\chi}
	\renewcommand*\d{\delta}
	\newcommand*\D{\Delta}
	\renewcommand*\l{\lambda}
	\renewcommand*\L{\Lambda}
	\renewcommand*\t{\tau}
	\renewcommand*\u{\upsilon}
}

\newcommand*\rcvr@tmp[1]{\edef\next{\global\let\noexpand#1\csname#1@tmp}\next}

\newcommand*\longGreek{
	\rcvr@tmp{a}
	\rcvr@tmp{b}
	\rcvr@tmp{c}
	\rcvr@tmp{d}
	\rcvr@tmp{D}
	\rcvr@tmp{l}
	\rcvr@tmp{L}
	\rcvr@tmp{t}
	\rcvr@tmp{u}
}
	
\shortGreek

\renewcommand*\({\left(}
\renewcommand*\){\right)}

\renewcommand*\[{\left[}
\renewcommand*\]{\right]}

\newcommand*\?{}

\newcommand*\abso[1]{\verticalbar#1\verticalbar} 

\newcommand*\adj{^{\dagger}}	

\newcommand*\alphahat{{\widehat{\alpha}}}	

\newcommand*\At[1]{{}_{\big\vert#1}} 
\newcommand*\at[1]{_{\vert#1}}

\renewcommand*\Bar[1]{{\overline{#1}}}	

\newcommand*\bhat{\unit{b}}	

\newcommand*\bigast{{\mathchoice{\asterisk}
	{\displaystyle\asterisk}
	{\textstyle\asterisk}
	{\scriptstyle\asterisk}
	}}

\newcommand*\Cpp{C$++$} 	

\newcommand*\cS{{\mathcal{S}}}
\newcommand*\cT{{\mathcal{T}}}	

\newcommand*\cZ{{\mathcal{Z}}}

\newcommand*\cs{c_{\rm s}} 	

\newcommand*\conj{^\bigast}	

\newcommand*\cross{\boldsymbol{\times}} 

\newcommand*\Dv{D_v}		

\newcommand*\DIA{direct-interaction approximation}

\newcommand*\DT{\Delta T}		

 \newcommand*\Ekern{\kern-0.12em}

\newcommand*\dT{\delta T}		

\newcommand*\degreeso{^\circ}		
\renewcommand*\degrees{\ifmmode\degreeso\else$\degreeso$\fi} 

\newcommand*\del{\partial}		

\newcommand*\delt{\partial_t}

\newcommand*\delx{\partial_x}

\newcommand*\dely{\partial_y}

 \newcommand*\fkern{\kern-0.125em}

\newcommand*\diel{{\mathcal{D}}} 	
\newcommand*\dielperp{\diel_\perp} 	

\newcommand*\Dirac[1]{\delta(#1)}

\newcommand*\EB{\vE\cross\vB}

\newcommand*\Epar{E_\parallel}		

\newcommand*\ehalf{^{1/2}}		
\newcommand*\eg{\LatinAIP{e.g.}}
\newcommand*\ehat{\unit{e}}		

\newcommand*\ep{\epsilon_{\Rm p}}

\newcommand*\ETAL{\Latin{et~al.}}	
\newcommand*\etal{\ETAL\kill@period}

\newcommand*\ETC{\Latin{etc.}}		
\newcommand*\etc{\ETC\kill@period}

\newcommand*\FM{F_{\Rm M}}		

\newcommand*\ft{{\widetilde f}}		

\newcommand*\GK{gyrokinetic}
\newcommand*\GKE{\GK\ equation}

\newcommand*\Go{G^{(0)}}		

\newcommand*\grad{\nabla}		
\newcommand*\gradpar{\grad_\parallel}	
\newcommand*\gradperp{\grad_{\fkern\perp}}	

\renewcommand*\Hat[1]{{\widehat{#1}}}	
\newcommand*\HM{Hasegawa--Mima}
\newcommand*\HMe{\HM\ equation}
\newcommand*\HME{\HMe}

\newcommand*\half{\case12}		

\newcommand*\Int{\int\!\?}		
\newcommand*\I[2]{\int_{#1}^{#2}\!\?}	
\newcommand*\INT{\I{-\infty}\infty}	

\renewcommand*\Im{\Mathop{Im}}		

\newcommand*\ie{\LatinAIP{i.e.}}	

\newcommand*\kD{k_{\Rm D}}		
\newcommand*\kDi{k_{{\Rm D}i}}		
\newcommand*\kDe{k_{{\Rm D}e}}		
\newcommand*\kDs{k_{{\Rm D}s}} 		

\newcommand*\kbar{{\Bar{k}}}		
\newcommand*\kc{\kappa_{\Rm c}}		
	
\newcommand*\khat{\unit{k}}		

\newcommand*\kill@period{\futurelet\nextchar\no@period}
\newcommand*\no@period{\ifx\nextchar.\skip@period\fi}	

\newcommand*\kpar{k_\parallel}		
\newcommand*\kperp{k_\perp}		
\newcommand*\kperpbar{{\Bar{k}}_\perp} 

\newcommand*\kx{k_x}			
\newcommand*\ky{k_y}			

\newcommand*\Latin{\textit} 
\newcommand*\LatinAIP{\textrm}	

\newcommand*\LT{L_T}		

\newcommand*\lD{\lambda_{\Rm D}}	
\newcommand*\lDe{\lambda_{{\Rm D}e}}	

\newcommand*\lhs{left-hand side}

\newcommand*\Mathop[1]{\mathop{\hbox{\rm #1}}\nolimits}

\newcommand*\Max{\mathop{\operator@font max}}	

\newcommand*\Min{\mathop{\operator@font min}}	

\renewcommand*\tensor{\textsf}		

\newcommand*\mA{\tensor{A}}

\newcommand*\mC{\tensor{C}}
\newcommand*\mD{\tensor{D}}

\newcommand*\mI{\tensor{I}}
\newcommand*\mone{\mI}			

\newcommand*\mM{\tensor{M}}

\newcommand*\mS{\tensor{S}}

\newcommand*\mU{\tensor{U}}

\newcommand*\m[1]{^{-#1}}		

\newcommand*\mi{m_i}

\renewcommand*\ni{n_i}			

\newcommand*\norm[1]{\mathopen{\parallel}#1\mathclose{\parallel}}

\newcommand*\nuii{\nu_{ii}}	

\newcommand*\nuhat{\Hat{\nu}}

\newcommand*\Order[1]{O(#1)} 		

\newcommand*\Pb[1]{\set{#1}}		

\newcommand*\phbar{{\Bar{\ph}}}		

\newcommand*\pht{\Tilde\ph}		

\newcommand*\psit{\Tilde{\psi}}         

\newcommand*\qs{q_s}			

\newcommand*\Rm[1]{#1}			

\renewcommand*\Re{\Mathop{Re}}		

\newcommand*\ri{\rho_i}
\newcommand*\rs{\rho_{\textrm{s}}}	

\newcommand*\rhs{right-hand side}

\newcommand*\set[1]{{\let|\mid \{#1\}}}	

\newcommand*\Sstate{steady state}
\newcommand*\sstate{steady-state}

\newcommand*\Tbar{{\Bar{T}}}

\newcommand*\Tr{^{\Rm T}} 		

\newcommand*\Te{T_e} 		
\newcommand*\Ti{T_i} 		
\newcommand*\Tt{{\Tilde T}} 	

\renewcommand*\Tilde{\widetilde}	

\newcommand*\Ut{\widetilde U}	

\newcommand*\unit[1]{{\widehat{\v{#1}}}}	

\newcommand*\up[1]{^{(#1)}}

\newcommand*\Vkern{\kern-0.1em} 

\renewcommand*\v[1]{{\bm{#1}}} 	

\newcommand*\vB{\v{B}} 		

\newcommand*\vC{\v{C}}
\newcommand*\vD{\v{D}} 		
\newcommand*\vE{\v{E}} 		

\newcommand*\vS{\v{S}}

\newcommand*\vf{\v{f}}
\newcommand*\vgrad{\v{\nabla}}

\newcommand*\vk{\v{k}}

\newcommand*\vkw{{\vk,\omega}}		

\newcommand*\vpar{v_\parallel}

\newcommand*\vperp{v_\perp}
\newcommand*\vpsi{\v{\psi}}		

\newcommand*\vq{\v{q}}
\newcommand*\vqhat{\Hat{\vq}}		

\newcommand*\vti{v_{{\Rm t}i}}	

\newcommand*\vV{\v{V}}		
\newcommand*\vVE{\vV_{\Vkern E}}	

\newcommand*\vVEt{\Tilde{\vV}_{\Vkern E}}

\newcommand*\vv{\v{v}}

\newcommand*\vx{\v{x}}		

\newcommand*\Vt{{\Tilde V}}	

\newcommand*\Wk{\Omega_\vk}		

\newcommand*\wci{\omega_{{\Rm c}i}}

\newcommand*\wpi{\omega_{{\Rm p}i}}

\newcommand*\wrt{with respect to}

\newcommand*\xbar{{\Bar{x}}}

\newcommand*\xhat{\unit{x}}
\newcommand*\yhat{\unit{y}}

\newcommand*\ZF{zonal flow}	

\newcommand*\Partial[1]{\Deriv\partial\fr{#1}}

\newcommand*\PartiaL[1]{\Deriv\partial\fR{#1}}

\newcommand*\Total[1]{\Deriv{\?d}\fr{#1}}

\newcommand*\frD{\frac} 

\newcommand*\fr[1]{\fro#1\fro} 
 
\newcommand*\fR[1]{\left(\fr{#1}\right)}	

\newcommand*\Casefr[2]{\frac{#1}{#2}}	
\newcommand*\Case{\Casefr}

\newcommand*\casefr[2]{\mathchoice{{\textstyle\frD{#1}{#2}}}%
	{{\textstyle\frD{#1}{#2}}}
	{{\scriptstyle\frD{#1}{#2}}}%
	{{\scriptscriptstyle\frD{#1}{#2}}}}
\renewcommand*\case{\casefr}

\newcommand*\Half{\Casefr12}	

\newcommand*\Fourth{\Casefr14}

	\renewcommand*\BE{\begin{equation}}
	\renewcommand*\EE{\end{equation}}
	\renewcommand*\BEA{\begin{eqnarray}}
	\renewcommand*\BAL[1][]{\BM[#1]\BA}
	\newcommand*\BALams[1][]{\BM[#1]\BAams}
\def\BALams{\@ifnextchar[\BALams@{\BALams@[]}}
\def\BALams@[#1]#2\EALams{\BM[#1]\BAams#2\EAams\EM}
	\renewcommand*\EM{\end{subequations}}
	\newcommand*\WT{\begin{widetext}}
	\newcommand*\NT{\end{widetext}}
	\renewcommand*\etal{\ETAL}	
	\renewcommand*\etc{\ETC}	
	\renewcommand*\fkern{\!}	
	\renewcommand*\Vkern{}		
	\renewcommand*\Unskip{}		

\catcode`\@=12

\newcommand*\avg[1]{\langle#1\rangle}
\renewcommand*\Partial[2]{\frac{\partial#1}{\partial#2}}
\renewcommand*\PartiaL[2]{\left(\frac{\partial#1}{\partial#2}\right)}
\renewcommand*\Total[2]{\frac{d#1}{d#2}}

\renewcommand*\fr{\frac}
\renewcommand*\fR[2]{\left(\frac{#1}{#2}\right)}

\let\asterisk*
\def\.{\cdot}

\makeatletter

\def\FIGURE{\@ifnextchar[\@FIGURE{\@FIGURE[1]}}

\def\@FIGURE[#1]#2#3{\begin{figure}[ht]
\includegraphics[width=#1\columnwidth]{#2.eps}
	    \caption{#3}
	    \edef\@currentlabel{\thefigure}
	    \label{Fg.#2}
\end{figure}}

\long\def\comment#1\endcomment{}

\let\Bar\overline
\let\Hat\widehat
\let\Tilde\widetilde

\def\<#1>{\avg{#1}}

\let\adjo\adj
\renewcommand*\adj{\Mathop{adj}}

\renewcommand*\Cpp{C_{\ph\ph}}
\newcommand*\CpT{C_{\ph T}}
\newcommand*\CTp{C_{T\ph}}
\newcommand*\CTT{C_{TT}}

\newcommand*\Cppo{C_{\ph\ph}\up0}
\newcommand*\CpTo{C_{\ph T}\up0}

\newcommand*\CTTo{C_{TT}\up0}

\newcommand*\Cppp{C_{\ph\ph}\up+}
\newcommand*\CpTp{C_{\ph T}\up+}
\newcommand*\CTpp{C_{T\ph}\up+}
\newcommand*\CTTp{C_{TT}\up+}

\newcommand*\Cppm{C_{\ph\ph}\up-}
\newcommand*\CpTm{C_{\ph T}\up-}
\newcommand*\CTpm{C_{T\ph}\up-}
\newcommand*\CTTm{C_{TT}\up-}

\newcommand*\crit{_{\bigast}}

\newcommand*\dCpp{\D C_{\ph\ph}}
\newcommand*\dCpT{\D C_{\ph T}}
\newcommand*\dCTp{\D C_{T\ph}}
\newcommand*\dCTT{\D C_{TT}}
\newcommand*\dU{\D U}
\renewcommand*\dT{\D T}

\newcommand*\Dpp{D_{\z\z}}
\newcommand*\DpT{D_{\z T}}

\newcommand*\DTT{D_{TT}}

\newcommand*\ebar{\Bar{\eps}}
\newcommand*\eGV{\epsilon_{\rm GV}}
\renewcommand*\ehat{\Hat{\epsilon}}
\newcommand*\equil{\up{\rm eq}}

\newcommand*\F{\xi}

\newcommand*\fchar{f}
\renewcommand*\FM{F_{\rm M}}
\newcommand*\FMi{F_{{\rm M}i}}
\newcommand*\Fs{\xi_\bigast}

\newcommand*\ftz{\ft_\zeta}
\newcommand*\ftT{\ft_T}
\newcommand*\fz{f_\zeta}
\newcommand*\fT{f_T}

\newcommand*\f[1]{\def\temp{#1}
\delta
\ifx\temp\fchar
 \fkern
\fi
#1}

\newcommand*\Goo{\Gamma_0}
\renewcommand*\Go{\Hat{\Gamma}_0}
\newcommand*\gradbar[1]{\Bar{\grad}_{#1}}
\newcommand*\gradbarp{\gradbar+}
\newcommand*\gradbarm{\gradbar-}

\renewcommand*\hbar[1]{\Bar{h}_{#1}}
\renewcommand*\hm{\hbar-}
\newcommand*\hp{\hbar+}

\renewcommand*\kc{\kappa_{\rm c}}
\renewcommand*\khat{\Hat{\kappa}}
\renewcommand*\kperpbar{\kbar}
\newcommand*\ks{\kappa_\bigast}

\renewcommand*\l{\Hat{\lambda}}
\let\lo\lambda

\newcommand*\mDhat{\Hat{\mD}}
\newcommand*\mMhat{\Hat{\mM}}
\newcommand*\mm{\tensor{m}}
\newcommand*\mnu{\mathsf{\nu}}
\newcommand*\mnuhat{\Hat{\mnu}}
\newcommand*\mV{\tensor{V}}
\newcommand*\mVbar{\Bar{\mV}}

\newcommand*\nubar{\Bar{\nu}}
\newcommand*\nuzo{\nu_\zeta}
\newcommand*\nuz{\nuhat_\zeta}
\newcommand*\nuTo{\nu_T}
\newcommand*\nuT{\nuhat_T}
\newcommand*\nubaro{\nubar}

\newcommand*\PARTIAL[3]{\PartiaL{#1}
  {#2\smash{\rlap{\;\;\;$#3$}}}\phantom{#3}}  
\renewcommand*\pht{\Tilde{\ph}}

\renewcommand*\qs{q_\bigast}

\renewcommand*\set[1]{\{#1\}}

\newcommand*\trace{\Mathop{tr}}
\renewcommand*\Tt{\Tilde{T}}

\renewcommand*\U[1]{U_{#1}}
\newcommand*\Up{\U+}
\newcommand*\Um{\U-}
\renewcommand*\Ut{\Tilde{U}}

\newcommand*\ve{\v{e}}
\newcommand*\vpsibar{\Bar{\vpsi}}
\newcommand*\vfbar{\Bar{\vf}}
\newcommand*\vpsit{\Tilde{\vpsi}}
\renewcommand*\vqhat{\Hat{\vq}}
\renewcommand*\vS{\boldsymbol{\cS}}
\newcommand*\vshat{\Hat{\v{s}}}
\newcommand*\vShat{\Hat{\vS}}
\renewcommand*\Vt{\Tilde{V}}
\renewcommand*\vVEt{\Tilde{\vV}_{\vE}}

\newcommand*\x[1]{x_{#1}}
\let\xio\xi
\newcommand*\xihat{\Hat{\xio}}

\renewcommand*\ZF{^{\rm Z}}
\newcommand*\zt{\Tilde{\zeta}}

\newcommand*\xis{\xihat_\bigast}

\begin{document}

\title{Zonostrophic instability driven by discrete particle noise}

\author{D. A. St-Onge and J. A. Krommes}

\affiliation{Plasma Physics Laboratory, Princeton University, MS 28, P.O. Box
  451, Princeton, NJ  08543--0451}

\begin{abstract}
The consequences of discrete particle noise for a system possessing a
possibly unstable collective mode are discussed.  It is argued that a
zonostrophic instability (of homogeneous turbulence to the formation of
zonal flows) occurs just below the threshold for linear instability.  The
scenario provides a new interpretation of the random forcing that is
ubiquitously invoked in stochastic models such as the second-order cumulant
expansion (CE2) or stochastic structural instability theory (SSST); neither
intrinsic turbulence nor coupling to extrinsic turbulence is required.  A
representative calculation of the zonostrophic neutral curve is made for a
simple two-field model of toroidal ion-temperature-gradient-driven modes.
To the extent that the damping of zonal flows is controlled by the ion--ion
collision rate, the point of zonostrophic instability is independent of
that rate.
\end{abstract}

\date{\today}

\maketitle

\section{Introduction}

The zonostrophic instability\cite{Srinivasan-Young} of a statistically
homogeneous steady state to inhomogeneous zonal flows figures importantly
in current research on the physics of zonal-flow generation and the
interaction of zonal flows with turbulence.  The general framework, called
statistical state dynamics by Farrell and Ioannou,\cite{Farrell_Jets} is
intrinsically a statistical formalism --- perturbations are made to a
statistical ensemble, not a particular realization.  Analytical treatment
must therefore involve some sort of stochastic model.  Considerable
attention has been given to low-order cumulant truncations, in particular
the CE2 (second-order cumulant
expansion) studied by Tobias and
Marston\cite{Tobias11,Tobias-Marston13,Marston_Jets}  
and the closely related (in some 
cases mathematically identical) SSST or S3T (stochastic structural
stability theory) introduced earlier by Farrell and
Ioannou.\cite{Farrell03_SSST,Farrell_Dw_Zf}  A brief introduction to these
topics is given in Sec.~6.3 of reference~\REF{JAK_tutorial}.  In
CE2 the so-called eddy--eddy 
nonlinearities are neglected; only the interactions between the turbulence
and the zonal flows are retained.  Generally the turbulence is driven by a
white-noise stochastic forcing~$\ft$, for which various interpretations
have been given.  Sometimes, especially in S3T papers, it is taken to
represent the effects of the missing eddy--eddy interactions.  In strict
CE2, it is instead taken to represent the effects of extrinsic fluctuations
--- \eg, 
baroclinic instabilities --- whose dynamics are not described by the
partial differential equation under study.  In the present paper we introduce
a variant of this latter interpretation; we attribute~$\ft$ to the discrete
particle noise of a weakly coupled many-body plasma.  We describe the
relationship of this viewpoint to Kadomtsev's classic discussion of the
transition from stable to unstable collective modes, and we illustrate with
a calculation of the neutral curve for the zonostrophic instability of a
simple model of the toroidal ion-temperature-gradient (ITG) instability.
New insights about the onset of the Dimits shift follow.

In Kadomtsev's famous review/monograph,\cite{Kadomtsev} he considered the
following equation for fluctuation intensity~$I$ (we have altered his notation
slightly): 
\BE
\Total{I}{t} = 2\g I - 2\a I^2 + 2F.
\eq{I_dot}
\EE
Here $\g$~is the linear growth rate of a collective mode, $\a$~is a
positive mode-coupling coefficient, the $I^2$ term describes the
possibility of nonlinear saturation of the linear instability, and
$F$~describes the (small) level of noise due to discrete particles.
(This equation and many other facets of Kadomtsev's book are discussed in
  a lengthy tutorial article by Krommes.\cite{JAK_tutorial})
In thermal equilibrium, $\g$~can be interpreted as the negative of the
Landau-damping rate of a typical fluctuation with $k\lD \ll 1$, where
$\lD$~is the Debye length [$\lD\m2 = \sum_s \kDs^2$, with $\kDs \doteq
(4\pi n_s q_s^2/T)\ehalf$ being the Debye wave number for species~$s$].
Out of thermal equilibrium, it is assumed that $F$~remains unchanged while
$\g$~changes as an
order parameter (\eg, the temperature gradient~$\k$) is varied.
For stable plasma ($\g < 0$), the \sstate\ balance is 
\BE
I \approx F/\abso{\g}.
\eq{I_q}
\EE
As $\g \to 0_-$, that approximate level diverges.  However, as the fluctuations
become sufficiently large, the mode-coupling term takes over and permits a
smooth transition through the point $\g = 0$; for large~$\g$, the nonlinear
saturation level is $I \approx \g/\a$.  Of course, the \sstate\
solution of the quadratic equation \EQ{I_dot} can be found exactly; it is
graphed in \Fig{I_ss}. 

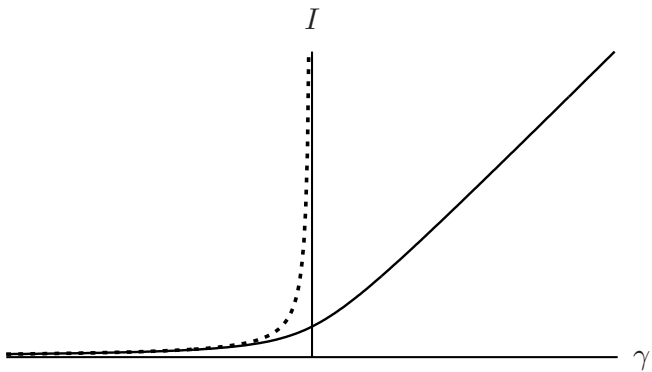
\FIGURE{I_ss}{Solid curve:  Steady-state solution of \Eq{I_dot}.  Dashed
  curve:  Approximate solution when the quadratic mode-coupling term is
  neglected.}

The mode-coupling term $\propto I^2$ in \Eq{I_dot} implicitly assumes
turbulence; it represents the net effect of the statistical
closure\cite{JAK_PR,JAK_tutorial} of a quadratic nonlinearity.  Thus it
  does not capture the Dimits-shift phenomenon.  The Dimits shift was first
  observed in the computer simulations described in \Ref{Dimits_shift}.  It
  refers to the fact that as the background temperature gradient~$\k$~is
  varied from below to above the linear 
  threshold~$\kc$ for instability (the focus of Dimits \etal\ was on the ITG
  instability), the ion heat flux~$Q(\k)$ turns on not for $\k > \kc$ but
  only for a larger value $\k > \ks > \kc$.  The difference $\ks - \kc$ is
  called the Dimits shift.  It is understood that the suppression of heat
  flux in the Dimits-shift regime is due to the excitation of zonal
  flows.\cite{Dimits_shift,Rogers00} In
  that regime, there is no 
  turbulence, so conventional statistical closure does not apply.  It is of
  interest to understand how the transition to turbulence is modified when
  the physics of the Dimits shift is included.  

In the parlance of CE2 or S3T, the mode-coupling considered by Kadomtsev
describes only the effect of the eddy--eddy interactions.  Those
stochastic models,
which although quite simple have been surprisingly successful in
various contexts,
 neglect the eddy--eddy interactions altogether, but do
consider the interaction of zonal modes with turbulence.  They are also
tractable, so they provide a good starting point for further
investigations.  However, in the regimes of linear stability and the Dimits
shift, there is no turbulence (we exclude the possibility of subcritical
turbulence in this discussion), so the S3T interpretation of~$\ft$ as
representing the effects of turbulent eddy--eddy interactions is not
viable.  Instead, we shall use CE2 and interpret~$\ft$ as being due to discrete
particle noise, which is an extrinsic forcing from the point of view of the
collective ITG dynamics.  We
assume that zonal modes are not driven directly by the particle noise (a
very weak effect), but only by the Reynolds stress due to collective
modes.  The following scenario then pertains.  In the absence of zonal
flows, the balance \EQ{I_q} between 
particle noise and modal damping creates a statistically homogeneous
\Sstate\ (in agreement with Kadomtsev's interpretation).  For sufficiently
large Landau damping or sufficiently small~$\k$, the homogeneous 
fluctuation level is small and the homogeneous state is stable against the
formation of inhomogeneous zonal modes (which suffer a small but finite
amount of collisional damping).  Now imagine increasing~$\k$,
thereby 
driving the collective modes toward the threshold for linear instability.
If zonal modes would never form, then 
the fluctuation intensity would diverge as $\g \to 0_-$, as does the dashed
curve in \Fig{I_ss}.  However, as $\g \to 0_-$ the forcing
represented by the \rhs\ of \Eq{I_q} becomes relatively large and, on the
average, 
overcomes the damping on the zonal modes; a zonostrophic
instability\cite{Srinivasan-Young} (a supercritical bifurcation)
occurs somewhere to the left of the linear threshold $\g = 0$ ($\k = \kc$).
These two 
thresholds are clearly distinct, with the zonostrophic 
one involving nonlinear effects.  In the context of the ITG problem,
the latter threshold defines the left-hand boundary of the Dimits-shift
regime.\cite{Dimits_shift}  This interpretation of that boundary, as being
related to a zonostrophic instability driven by discrete particle noise, is
new.  

To the extent that the 
particle noise is very small (as it is in hot, magnetically confined fusion
plasmas), it might be thought that the zonostrophic
threshold is very 
close to the linear one and that in the collisionless
limit where the noise approaches zero the two
thresholds should become 
coincident.  Thus in neither the discussion by Dimits \etal\ in
\Ref{Dimits_shift} 
of collisionless simulations nor the earlier work on the ITG Dimits shift
by Kolesnikov and Krommes\cite{Kolesnikov_PRL,Kolesnikov_PoP}
is there any mention of particle noise.  
In fact, however,
a basic scaling with the discreteness parameter~$\ep \doteq (n\lD^3)\m1$
cancels out in the 
competition between the forcing and the 
zonal damping that defines the zonostrophic transition
because the zonal damping rate is proportional to the ion--ion
collision rate~$\nuii$, which of course scales with the discrete-ion noise
period, 
  However, in \GK
s\footnote{An introductory review article on \GK s that contains references
  to more 
  specialized reviews and original papers is by J.~A. Krommes, Annu.\ Ref.\
  Fluid Mech.\textbf{44}, 175 (2012).}
the equilibrium fluctuation level\cite{JAK_es,Nevins_noise}
 also scales with the inverse of the large
perpendicular dielectric function $\dielperp \doteq \wpi^2/\wci^2 =
\rs^2/\lDe^2 \gg 1$.  [Here $\wpi \doteq (4\pi \ni q_i^2/\mi)\ehalf$ is
the ion plasma frequency,
$\wci \doteq q_i B/\mi c$ is the ion gyrofrequency, 
$\lDe \doteq \kDe\m1$,
and $\rs \doteq \cs/\wci$ is the sound radius, where $\cs \doteq
(Z\Te/\mi)\ehalf$ is the sound speed.]
($\dielperp$~arises because of the shielding effect due to ion
polarization.)  Because $\nuii$~does not involve~$\dielperp$, the distance
of the 
zonostrophic threshold to the linear threshold scales
with~$\dielperp\m1$.  That is also small, but not nearly as small as the
typical discreteness parameter~$\ep$.

In the remainder of this paper we shall make this idea quantitative by
showing how to calculate the neutral curve (the marginality condition for
the onset of zonostrophic instability as a function of zonal wave number~$q$)
associated with an ultra-simple
two-field fluid model of the toroidal ITG mode.  The purpose of the analysis is
primarily to illustrate conceptual principles and to establish the
consistency of the 
interpretation.  We do not attempt to incorporate all details of the noise
sources in the equations for fluctuating potential and temperature (a
proper calculation should be kinetic, whereas we use a fluid description),
so we 
cannot be fully quantitative --- and of course the model itself lacks many
details of importance in practice.  Nevertheless, the calculation
demonstrates the basic 
concept of the onset of the noise-driven zonostrophic instability, it
shows how to extend to a two-field model the earlier one-field 
analyses of Srinivasan and Young\cite{Srinivasan-Young} and Parker and
Krommes,\cite{Parker-JAK_jets_PoP,Parker-JAK_NJP,Parker-JAK_Jets} 
and it makes a prediction for the wave number of the first zonal mode~$\qs$
that is driven unstable.

\comment
\emph{The idea would be to create a figure analogous to \Fig{I_ss} showing
  the intensities of the drift waves and zonal flows as a function
  of~$\g$, from $\g < 0$ to partway into the Dimits-shift regime.}
\endcomment

\section{Two-field model of the ion-temperature-gradient-driven mode}

The ion-temperature-gradient-driven (ITG) mode is believed to be
responsible for the anomalously large ion heat losses in the cores of
modern tokamaks.   It has both a slab branch and a toroidal
branch,\cite{Chen-Cheng} and it has been extensively studied both
numerically\cite{Ottaviani_2field,Hammett_developments,GYRO_PRL}
and analytically.\cite{Biglari_toroidal_ITG,Romanelli_Briguglio,Ottaviani_ITP,Rogers00}
In this paper we adopt the simplest possible model that possesses a
toroidal ITG mode.  Thus we consider a two-dimensional two-field gyrofluid
model that retains only the curvature drift, an ion temperature gradient
(with a flat density profile), and the advective $\EB$ nonlinearity.  We
use the usual plasma slab coordinates in which $x$ and~$y$ correspond to the
radial and poloidal directions, respectively; we neglect all parallel ($z$)
dynamics.  With time and space being scaled to~$a/\cs$ and~$\rs$,
respectively ($a$~is the minor radius), 
 the equations are\cite{Ottaviani_ITP}
\BALams[zT]
\delt\zt + \vVEt\.\vgrad\zt + \e\,\dely\Tt &= -\nuzo\zt + \ftz,
\\
\delt\Tt + \vVEt\.\vgrad\Tt + \k\,\dely\pht &= -\nuTo\Tt + \ftT.
\EALams
Here tilde denotes a random variable;
$\ph \doteq (e\p/\Te)(a/\rs)$ is the dimensionless electrostatic potential;
the definition and properties of the $\EB$ velocity are
\BALams
&\vVEt \doteq \bhat\cross \vgrad\pht = -\dely\pht\, \xhat + \delx\pht\, \yhat
\equiv \Ut\,\xhat + \Vt\,\yhat,&
\\
&\vVE\.\vgrad A = \Pb{\ph,A} \equiv (\delx \ph)(\dely A) - (\dely
\ph)(\delx A)\rlap{;}& 
\EALams
the generalized vorticity is
\BE
\z \doteq (\gradperp^2  - \alphahat)\ph \equiv \gradbar{}^2\ph,
\EE
where $\alphahat$~is an operator that vanishes when acting on zonal modes
and is unity otherwise;
$\e \doteq 2a /R$ describes the curvature drive;
$\k \doteq a/\LT$, where $\LT \doteq -d\ln\Ti/d\ln x$ is taken to be
constant; 
$\Tt \doteq (\Tt_i/\Te)(a /\rs)$ ($\Te$~is taken to be constant) describes the
deviation of the ion temperature profile from one with constant~$\k$;
$\nuzo$ and~$\nuTo$ represent damping operators that in Fourier space are
assumed to be even in both~$\kx$ and~$\ky$; 
and the~$\ft$'s represent the random particle noise.  A significant
approximation is to neglect finite-Larmor-radius (FLR) effects; this
precludes detailed comparisons with the equations and results of
\Ref{Rogers00}. In the absence of forcing, damping, and zonal modes, the
linearized system
\BE
\delt\D\z + \e\,\dely\DT = 0,
\quad
\delt\DT + \k\,\dely\D\ph = 0
\EE
implies the dispersion relation 
\BE
\lo^2 = \Goo^2
\EE
(time variations $e^{\lo t}$ are assumed), where
\BE
\Goo \doteq \ky\sqrt{\k\e}/\kperpbar
\eq{Go}
\EE
and
\BE
\kperpbar^2 \doteq 1 + \kperp^2.
\EE
The eigenvalue $\lo_+ \approx \Goo$, with unnormalized eigenvector $\ve \doteq
(1,\,-i\sqrt{\k/\e}/\kperpbar)\Tr$, is the unstable toroidal ITG mode.
With small dissipation added, the eigenvalues are
\BE
\lo_\pm = \pm[\Goo^2 + (\nuzo - \nuTo)^2/4]\ehalf - \nubaro \approx \pm\Goo -
\nubaro,
\EE
where
\BE
\nubaro \doteq \Half(\nuzo + \nuTo)
\eq{nubar_def}
\EE
and the approximation holds for $\Goo \gg \nubaro$.  With
\BE
\xio \doteq \Goo^2 - \nuzo\nuTo,
\eq{G2}
\EE
the transition point $\lo = 0$ corresponds to $\xio = 0$, with
$\xio > 0$ defining the regime of linear instability.  

It will be shown in \Sec{constraints} that consistency of the model
requires that $\nuzo = \nuTo = \nubaro = \nu$, where $\nu$~can be interpreted
as the 
Landau-damping rate of the electrostatic potential.  Let quantities
normalized to~$\nu$ be denoted by a hat, \eg,
\BE
\Hat{\k} \doteq \k/\nu,
\quad
\Go \doteq \ky\sqrt{\Hat{\k}\Hat{\e}}/\kperpbar,
\quad
\xihat \doteq \xio/\nu^2,
\quad
\hbox{etc.};
\EE
then the regime of linear stability is $-1 \le \xihat \le 0$.
For now, however, we
shall keep~$\nuzo$ and~$\nuTo$ distinct in order to help keep track of the
origin of various terms.

\let\nuhat\nu

\section{The second-order cumulant expansion (CE2)}

\subsection{General strategy}

We shall treat \Eqs{zT} by means of the stochastic model known as the CE2
(second-order cumulant
expansion).\cite{Tobias11,Tobias-Marston13,Marston_Jets}    The basic
strategy is to decompose the 
fields into mean and fluctuating parts, \eg, $\Tt = \Tbar + \f T$, where the
overline denotes a zonal average,
then to ignore products of fluctuating terms (the ``eddy--eddy''
interactions).  An ergodicity assumption is also made, so the barring
operating is taken to be 
equivalent to the ensemble average $\<\dots>$ over the microscopic state;
we assume that $\<\ft> = 0$.  
The resulting system that couples the mean and fluctuating fields is known
as the 
quasilinear approximation.  Without further approximation, it can be closed
exactly by constructing equations for the two-point space-time correlation
functions.  Because the statistics are constructed from primitive amplitude
equations, they are guaranteed to be realizable, \ie, to be compatible with
a legitimate probability density functional.  For example, the solution of
the equations for the two-point
correlation matrix is guaranteed to be a positive-semidefinite form.  For
an introduction to realizability and statistical closure in this context,
see \Ref{JAK-Parker_Jets}. 

Although it is possible to close at the level of two-time-point correlation
functions, 
generally closure is done at the level of one-time
correlation functions by assuming that the~$\ft$'s are white noise (delta
correlated in time).  The resulting equations, which define the standard
CE2 approximation and will be written below,
are also realizable.

Use of a white-noise approximation may be questionable, since clearly the
physical fluctuations are not white.  However, this method has a successful
track record, not only in the present CE2 context but also in the general
theory of statistical closures.\cite{JAK_PR}  Thus the \DIA\
(DIA),\cite{Kr59} which is nonlocal in time and has a nontrivial
representation in frequency space,\cite{Kadomtsev} has a Langevin
representation\cite{Leith71,Kr_convergents} in which the forcing is not
white.  However, a 
related Markovian approximation\cite{TFM,JAK-Parker_Jets} does use white
forcing.  While such 
an approximation cannot do complete justice to the details of two-time
correlations, it has been shown to be reasonably successful at predicting
single-time wave-number spectra.  We view the CE2 in the same light.

\subsection{CE2 equations for the ITG model}

For the 2D ITG model, we define
the zonal mean of an arbitrary quantity~$A(\vx)$ as\footnote{In this paper
we use the standard plasma-physics slab coordinate system in which
$x$~represents the direction of inhomogeneity (the radial coordinate in a
tokamak) and $y$~represents the zonal direction (the poloidal direction in
a tokamak).  (In detail, in an actual tokamak the zonal flows are in the
direction perpendicular to both~$x$ and~$\vB$, \ie, mostly in the poloidal
direction for large aspect ratio.)  In geophysics, the roles of~$x$ and~$y$
are interchanged.}
\BE
\Bar{A}(x) \doteq \fr{1}{L_y}\I0{L_y} dy\,A(x,y).
\EE
Upon dropping the overlines on the zonally
averaged fields, the equations for the mean fields are
\BALams[means]
\delt U(x,t) &= -\delx(\Bar{\f u\,\f v}) - \nuz\ZF U,
\eq{U_dot}
\\
\delt T(x,t) &= -\delx(\Bar{\f u\,\f T}) - \nuT\ZF T
\eq{T_dot}
\EALams
where $U(x,t) \doteq \delx\phbar$ is the $y$-directed zonal velocity.
In arriving at \Eq{U_dot}, we integrated the equation for~$\Bar{\z}$ once
in~$x$.  The fluctuations obey
\BALams[fluctuations]
\delt\f\z &= -U\,\dely\f\z + (\delx^2 U)\dely\f\ph - \e\,\dely\f T - \nuz
\,\f\z + 
\f \fz,
\\
\delt \f T &= -U\,\dely \f T - (\k - \delx T)\dely\f \ph - \nuT\, \f T + \f\fT.
\EALams
To construct the CE2 equations, we define the two-space-point covariance
tensor 
\BAams
&C_{AB}(\x1,\x2,t;\ky) 
\NN\\
&\ \doteq \INT
dr\,e^{-i\ky r}\avg{\f A(\x1,y_2+r,t)\f B(\x2,y_2,t)}; 
\EAams
this quantity is independent of~$y_2$ by virtue of statistical homogeneity
(translational invariance) in~$y$.  We shall drop the $\ky$~argument when
there is no possibility of confusion.  Similarly, we assume that the
forcing is stationary, homogeneous white noise and write
\BALams
&\INT dr\,e^{-i\ky r}\<\f f_A(\vx_1,t)\f f_B(\vx_2,t')> 
\NN\\
&\qquad \doteq F_{AB}(\x1,t;\x2,t';\ky)
\NN\\
&\qquad = 2D_{AB}(\x1-\x2;\ky)\Dirac{t-t'}.
\EALams
In terms of $U_i \doteq U(x_i)$, $T_i \doteq T(x_i)$, $\gradbar i^2 \doteq
\del_{x_i}^2 - 
\ky^2 - 1$, and $\D_i \doteq \k - \del_{x_i}T_i$,
one then finds
\begin{widetext}
\BALams[C12]
\delt\gradbar1^2\gradbar2^2 \Cpp
&= -i\ky(\U1 - \U2)\gradbar1^2\gradbar2^2\Cpp
+ i\ky(\U1''\gradbar2^2 - \U2''\gradbar1^2)\Cpp
- i\ky \e(\gradbar2^2\CTp - \gradbar1^2 \CpT)
\NN\\
&\quad
- 2\nuz\gradbar1^2\gradbar2^2\Cpp + 2\Dpp,
\\
\delt\gradbar1^2\CpT &= -i\ky[(\U1 - \U2)\gradbar1^2 -
  (\del_{\x1}^2\U1)]\CpT + i\ky \D_2\gradbar1^2\Cpp
- i\ky\e\CTT
- 2\nubar\gradbar1^2\CpT
+ 2\DpT,
\\
\delt\CTT &= -i\ky(\U1 - \U2)\CTT
- i\ky(\D_1\CpT - \D_2\CTp)
- 2\nuT\CTT + 2\DTT,
\EALams
\end{widetext}
together with $\CTp(\x1,\x2) = \CpT\conj(\x2,\x1)$.

More informative coordinates are the sum and difference variables
\BE
\xbar \doteq \Half(\x1 + \x2),
\quad
x \doteq \x1 - \x2,
\EE
which isolate
any dependence on statistical inhomogeneity in~$\xbar$.  The inverse
transformation is
\BE
\x1 = \xbar + \Half x,
\quad
\x2 = \xbar - \Half x.
\EE
The use of~$x$ and~$\xbar$ enable one to make contact with the theory of
Wigner--Moyal transforms.\cite{Ruiz-Parker}
Also define the modified Laplacian
\BE
\gradbar\pm^2 \doteq \gradbar x^2 \pm \del_x \del_{\xbar} +
\Fourth\del_\xbar^2,
\EE
where $\gradbar x^2 \doteq \del_x^2 - \ky^2 - 1$, as well as
\BE
\U\pm \doteq U\(\xbar \pm \half x\),
\EE
and write $C(\x1,\x2,\ky,t) \equiv C(x,\ky \mid \xbar,t)$.
The transcription of \Eqs{C12} is then immediate:
\begin{widetext}
\BALams[Cpm_dots]
\delt\gradbarp^2\gradbarm^2\Cpp &= -i\ky(\Up - \Um)\gradbarp^2\gradbarm^2\Cpp
+ i\ky(\Up''\gradbarm^2 - \Um''\gradbarp^2)\Cpp
- i\ky \e(\gradbarm^2\CTp - \gradbarp^2\CpT)
\NN\\
&\quad - 2\nuz\gradbarp^2\gradbarm^2\Cpp
+ 2\Dpp,
\\
\delt\gradbarp^2\CpT &= -i\ky[(\Up - \Um)\gradbarp^2 - \Up'']\CpT
+ i\ky\D_-\gradbarp^2\Cpp - i\ky\e\CTT
- 2\nubar\gradbarp^2\CpT
+ 2\DpT,
\\
\delt\CTT &= -i\ky(\Up - \Um)\CTT
- i\ky(\D_+ \CpT - \D_-\CTp)
- 2\nuT\CTT + 2\DTT,
\EALams
\end{widetext}
together with $\CTp(x\mid\xbar) = \CpT\conj(-x\mid\xbar) = \CpT(x \mid
-\xbar)$.
The mean equations are
\BALams[means_dot]
\(\Partial{}{t} + \nuz\)U(\xbar,t) &= -i\Partial{}{\xbar}\Int\fr{d\ky}{2\pi}\,
\ky(\delx\Cpp)\at{x=0},
\eq{UZ_dot}
\\
\(\Partial{}{t} + \nuT\)T(\xbar,t) &=
\Half i\Partial{}{\xbar}\Int\fr{d\ky}{2\pi}\,\ky(\CpT - \CTp)\at{x=0}.
\eq{TZ_dot}
\EALams
In the derivation of \Eq{UZ_dot}, a term $\propto \ky\del_{\xbar}\Cpp(0,\ky
\mid \xbar,t)$ vanished by symmetry under the $\ky$~integration.  The
result \EQ{TZ_dot} has been written in a convenient symmetrized form.

\section{CE2 analysis of the ITG model}

\subsection{Homogeneous steady states of the ITG model}

We denote 
statistically steady ($\delt = 0$) and homogeneous ($\del_\xbar = 0$)
solutions by a superscript~(0).
One obtains $U\up0 = T\up0 = 0$, which permits a 
ready Fourier transformation in~$x$.  Thus $\gradbar\pm^2 \to
-\kperpbar^2$, where $\kperpbar^2 \doteq 1 + \kperp^2$, and
the components of the equilibrium covariance matrix
$\mC\up0$ obey
\BALams
0 &= 2\ky\e\kperpbar^2\Im\CpTo - 2\nuz\kperpbar^4\Cppo + 2\Dpp,
\eq{Cpp0}
\\
0 &= -i\ky\k\kperpbar^2\Cppo -i\ky\e\CTTo
\NN\\
&\quad + 2\nubar\kperpbar^2\CpTo + 2\DpT,
\eq{CpT0}
\\
0 &= 2\ky\k\Im\CpTo - 2\nuT\CTTo + 2\DTT,
\eq{CTT0}
\EALams
The real part of $\CpTo$ follows from the real part of \Eq{CpT0}:
\BE
\Re\CpTo \equiv \CpTo{}' = -\DpT'/\nubar\kperpbar^2,
\eq{CpT0'}
\EE
\Equations{Cpp0}, \EQ{CTT0}, and
the imaginary part of \Eq{CpT0} then define a 3D linear algebraic system
that can be solved for~$\Cppo$, $\Im\CpTo \equiv \CpTo{}''$, and~$\CTTo$ in
terms of the as-yet-unspecified noise sources.
The algebra is tractable by hand; one finds
\def\nuhat{\Hat{\nu}}
\begin{widetext}
\BE
\begin{pmatrix}
\Cppo \\
\CpTo{}''\\
\CTTo
\end{pmatrix}
 = \fr{1}{2\nubar\kperpbar^4\xi}
\begin{pmatrix}
\xi - \nuTo^2 & \ky\e\nuTo & -\ky^2\e^2\\
-\ky\k\nuTo & \kperpbar^2\nuzo\nuTo & -\ky\kperpbar^2\e\nuzo\\
-\ky^2\k^2 & \ky\k\kperpbar^2\nuzo &
\kperpbar^4(\xi - \nuzo^2)
\end{pmatrix}
\begin{pmatrix}
\Dpp\\
2\DpT''\\
\DTT
\end{pmatrix}
\to \fr{1}{2\kperpbar^4\xihat}
\begin{pmatrix}
\xihat - 1 & \ky\ehat & -\ky^2\ehat^2\\
-\ky\khat & \kperpbar^2 & -\ky\kperpbar^2\ehat\\
-\ky^2\khat^2 & \ky\khat\kperpbar^2 & \kperpbar^4(\xihat - 1)
\end{pmatrix}
\fr{1}{\nubar}\begin{pmatrix}
\Dpp\\
2\DpT''\\
\DTT
\end{pmatrix},
\eq{mC0}
\EE
\end{widetext}
where the last result follows by setting all of the~$\nu$'s equal.
We shall show in \Sec{constraints} that for discrete particle noise $\mDhat =
\nubar\mC\equil$ [\Eq{mD_eq}], so \Eq{mC0} describes the amplification of
the 
equilibrium fluctuations by the factor~$\abso{\xihat}\m1$, where $\xihat
\to 0_-$ 
as the linear threshold is approached.  As a check, when $\khat = \ehat =
0$ ($\xihat = -1$), the matrix in \Eq{mC0} becomes diagonal and correctly
recovers the equilibrium~$C$'s (with the~$\kperpbar^2$ converting
between~$\ph$ and~$\z$).

We should verify that our formula for~$\mC\up0$ is realizable.
We shall be interested in forcing such that $\DpT'' = 0$.  Then, since $\xi
< 0$ defines the regime of linear stability, one can see that the diagonal
elements $\Cppo$ and~$\CTTo$ are realizable (positive) up to the linear
threshold where \hbox{$\xi = 0$}. At that point all of quantities on the \lhs\ of
\Eq{mC0} 
diverge to infinity.  One must also check that $\mC$~is a positive-definite
form for all realizable forcings.  This is shown in \Sec{detC} to be true
up to the linear threshold.

\subsection{Zonostrophic instability of the ITG model}

\subsubsection{The general dispersion relation}

To determine the stability of the homogeneous equilibrium, we add to each
equilibrium quantity~$Q\up0$ a perturbation of the form $\d Q e^{\l
  t}e^{iq\xbar}$, linearize \Eqs{Cpm_dots} and (the already linear)
\Eqs{means_dot}, Fourier transform in the difference variable~$x$, and
ultimately derive a dispersion relation. Define  
\BE
\hbar\pm^2 \doteq \(\kx \pm \Half q\)^2 + \ky^2 + 1 = \kperpbar^2 \pm \kx q +
\Fourth q^2 > 0. 
\EE
Then upon defining $C\up\pm \doteq C\up0(\kx \pm \half q,\ky)$,
the perturbed equations become
\begin{widetext}
\BALams[dCs]
(\lo + 2\nuzo)\hbar+^2\hbar-^2\dCpp &= -i\ky[\hbar-^2(\hbar-^2 - q^2)\Cppm
  - \hbar+^2(\hbar+^2 - q^2)\Cppp]\dU
- i\ky\e(\hbar+^2\dCpT - \hbar-^2\dCTp),
\\
-(\lo + 2\nubar)\hbar+^2\dCpT &= i\ky[(\hbar-^2 - q^2)\CpTm -
  \hbar+^2\CpTp]\dU
- i\ky\k\hbar+^2\dCpp 
- \ky q\hbar+^2\Cppp\dT
- i\ky \e\,\dCTT,
\\
-(\lo + 2\nubar)\hbar-^2\dCTp &= -i\ky[(\hbar+^2 - q^2)\CTpp -
  \hbar-^2\CTpm]\dU
+ i\ky\k\hbar-^2\dCpp 
+ \ky q\hbar-^2\Cppm\dT
+ i\ky \e\,\dCTT,
\\
(\lo + 2\nuTo)\dCTT &= -i
\ky(\CTTm - \CTTp)\dU
-i\ky \k(\dCpT - \dCTp)
- \ky q(\CpTm - \CTpp)\dT.
\EALams
\end{widetext}
If the perturbed variances are arranged as the column vector $\D\vC \doteq
(\dCpp,\,\dCpT,\,\dCTp,\,\dCTT)\Tr$, then after dividing each of \Eqs{dCs}
by~$\nubar$ one can write the system \EQ{dCs} as
\BE
\mMhat\.\D\vC = \vshat_U\dU + \vshat_T\dT,
\eq{MvC}
\EE
where the elements of~$\mMhat$, $\vshat_U$, and~$\vshat_T$ are easily identified
from \Eqs{dCs}.  (Again, the hats denote normalization \wrt~$\nubar$.)
The solution of \Eq{MvC},
\BE
\D\vC = (\mMhat\m1\.\vshat_U)\dU + (\mMhat\m1\.\vshat_T)\dT,
\EE
 then provides the components needed 
to evaluate the Reynolds stresses in the perturbed \Eqs{means_dot}, which
can be written as
\BALams[dmeans_dot]
(\lo + \nuzo\ZF)\dU &= iq\Int \fr{d\vk}{(2\pi)^2}\,
\kx\ky \dCpp,\\
(\lo + \nuTo\ZF)\dT &= -\Half q\Int \fr{d\vk}{(2\pi)^2}\,\ky(\dCpT - \dCTp).
\EALams
From \Eq{MvC}, the \rhs s of \Eqs{dmeans_dot} can be written as the
negative of a $2\times2$ matrix~$\mm$ operating on
$(\dU,\,\dT)\Tr$.
The dispersion relation is then 
\BE
\det(\mA) = 0,
\eq{disp}
\EE
where
\BE
\mA \doteq \lo\mone + \mnu\ZF + \mm(\lo)
\EE
and $\mnu\ZF = \Mathop{diag}(\nuzo\ZF,\,\nuTo\ZF)$.  The details of~$\mm$ are
recorded in \App{Dispersion}.   

We now introduce the concept of the neutral curve, which for fixed zonal
damping describes the forcing strength for which the zonostrophic growth
rate vanishes.  
In the simpler contexts of the barotropic vorticity
equation\cite{Srinivasan-Young} and the generalized
\HME,\cite{Parker-JAK_jets_PoP,Parker-JAK_NJP,Parker-JAK_Jets} the
bifurcation is known to be of Type~I$_{\rm s}$ in the language of pattern
formation,\cite{Cross-Greenside}
 meaning that the onset of instability occurs at nonzero~$q$ and
$\Im\l = 0$.  The same type of bifurcation can be shown to obtain in the
present problem, and a cartoon of such a neutral curve is shown in
\Fig{neutral_curve}.  
Note that in those earlier calculations with a scalar field, a single
dimensionless parameter describes 
the effective forcing.  Here we have multiple zonal dampings, but one can
introduce a common scaling parameter~$\eta$ that multiplies both~$\nuzo\ZF$
and~$\nuTo\ZF$ and sets their nominal sizes.  Furthermore, the forcing
matrix has a common scaling with 
the level~$\eps$ of particle noise.  We hold that level fixed (the relationship
between the various elements of~$\mD$ will be discussed below).  
Then the effective forcing is controlled by the modal instability
parameter~$\xi$ [\Eq{G2}]. 
Because the level of the homogeneous equilibrium diverges to~$\infty$ as
$\xi \to 0_-$, it is intuitively clear, and will be made more precise
below, that the zonostrophic bifurcation 
will occur just to the left of the linear stability threshold.  Given the
crude nature of the model, quantitative precision is unimportant, although
the calculation can certainly be done numerically and we shall display a
representative neutral curve in \Sec{Neutral}.  More importantly, it is of
interest to understand in principle how to calculate the bifurcation point
and the value of the zonal wave number at onset.

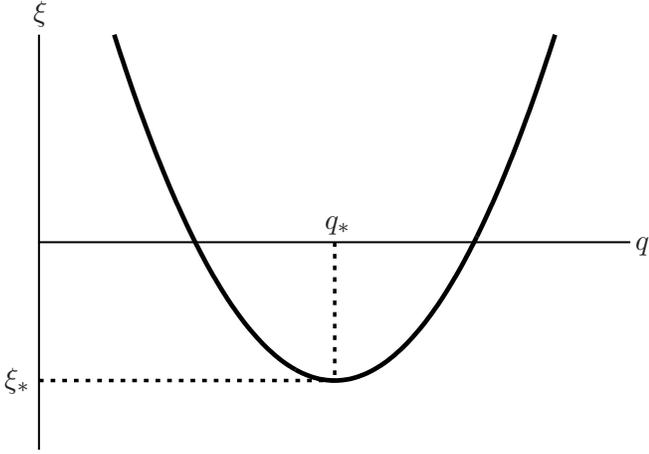
\FIGURE{neutral_curve}{A cartoon of a typical neutral curve, showing the
 bifurcation 
  point $(\qs,\,\Fs)$.  For the ITG model, $\xi = \G^2$ [\Eq{G2}], where
  $\xi = 0$ is the linear threshold.  For \GK\ particle noise, $\xis < 0$
  with $\abso{\xis} = \Order{\dielperp\m1} \ll 1$.}

To be more precise, observe from \Fig{neutral_curve} that at the point of
bifurcation the neutral curve~N has a minimum at a point $(\qs,\Fs)$.  By
definition, N~obeys
\BE
\l(q,\F) = 0,
\eq{l0}
\EE
which defines~N as a function $\F(q)$.  From
\BE
\PARTIAL{\l}{q}{\F}\,dq + \PARTIAL{\l}{\F}{q}\,d\F = 0,
\EE
one finds that on~N one has
\BE
\Total{\F}{q} = -\fr{(\del\l/\del q)\at{\F}}{(\del\l/\del \F)\at{q}},
\EE
so at the bifurcation point where $N$~has a minimum one has 
\BE
\PARTIAL{\l}{q}{\F} = 0.
\eq{lp0}
\EE
This
condition and \Eq{l0} provide two simultaneous equations to be solved
for~$\qs$ and~$\Fs$.  \Equation{l0} is just
\BE
\det[\mnu\ZF + \mm(0)] = 0,
\eq{det0}
\EE
where $\mm(0) \equiv \mm(\l{=}0,q,\xi)$.
To simplify the condition \EQ{lp0}, one can appeal to the Jacobi formula
for the derivative of the determinant of a matrix~$\mA$:
\BE
d\det(\mA) = \trace[\adj(\mA)\.d\mA],
\EE
where $\trace$~denotes the trace and $\adj$~denotes the adjugate, \ie, the
transpose of the cofactor 
matrix.  Upon differentiating \Eq{disp} and enforcing \Eq{lp0}, one obtains
\BE
0 = \trace\{\adj[\mnu\ZF + \mm(0)]\.\del q\,\mm(0)\}.
\eq{det'0}
\EE
Solution of the simultaneous nonlinear equations \Eq{det0} and \EQ{det'0}
determine the bifurcation point $(\qs,\,\Fs)$.

The elements of~$\mm$ are
linear functions of the elements of~$\mC$, and each of those elements is a
linear function of the elements of the noise matrix~$\mD$, which we take
to scale with the common level parameter~$\eps$:  $\mD = \eps\nubar\mDhat$,
where 
$\mDhat = \Order{1}$.  Also let $\mnu\ZF =
\eta\nubar\mnuhat\ZF$, where $\eta$~sets the common size of~$\nuzo\ZF$
and~$\nuTo\ZF$ 
such that $\mnuhat\ZF = \Order{1}$.
Since the elements of~$\mm$ are linearly proportional to those of~$\mD$,
it can be seen that the solution will depend only on the ratio $\ebar \doteq
\eps/\eta$.
Upon factoring~$\eta^n$
from each of \Eqs{det0} and \EQ{det'0}, where $n = 2$ for the ITG model,
those equations become 
\BALams
0 &= \det[\mnuhat\ZF + \mm(0,\qs,\xis;\ebar,\mDhat)] ,
\\
0 &= \trace\{\adj[\mnuhat\ZF +
  \mm(0,\qs,\xis;\ebar,\mDhat)]\.\del_{\qs}\mm(0,\qs,\xis;\ebar,\mDhat). 
\EALams
From \Eq{mC0}, we see that near linear threshold the divergent elements
of~$\mC\up0$ scale as $\mD/\nubar\xihat = \eta(\ebar/\xihat)\mDhat$. 
Therefore, one has
\BE
\mm(0,\qs,\xis;\ebar,\mDhat) \approx \mm(0,\qs,\xis/\ebar;\mDhat).
\EE
In the absence of additional small parameters in~$\mm$,
it is then clear that
the solution for $(\qs,\,\xis/\ebar)$ is $\Order{1}$, from which it follows
that $\xis = \Order{\ebar}$.  In cases in 
 which the forcing is due to turbulence and the zonal damping is weak,
 $\ebar$~is large.   Some analytical progress can be made by using
 isotropic ring forcing,\cite{Srinivasan-Young}
 $\mDhat(\kx,\ky) \propto \Dirac{k - k_{\rm f}}$.
 (Parker\cite{Parker-JAK_Jets} has used such forcing to show how the
 zonostrophic instability is a generalization of the modulational instability.)
 However, the situation is different when the forcing is
 due to discrete particle noise, as discussed next.

\subsubsection{Forcing due to particle discreteness}

In a tokamak, the zonal damping rates are expected to be proportional to
the ion--ion collision rate\cite{Rosenbluth-Hinton}:  $\eta \sim \nuii$.
Built into that rate is the noise level~$\e$ due to ion discreteness.
Therefore \emph{$\ebar$ is independent of the basic noise level,
  specifically the plasma parameter~$\ep$}.  However, it has been
shown\cite{JAK_es,Nevins_noise} 
that in magnetized, \GK\ plasma the thermal fluctuation level is reduced
by the strong shielding effect of ion polarization, \ie, that level scales
with the inverse of the perpendicular dielectric constant~$\dielperp$.
Thus $\ebar = \Order{\dielperp\m1} \ll 1$.  The particle-noise-driven
zonostrophic bifurcation therefore occurs at
\BE
\Go^2 = \nuz\nuT - a\ebar,
\EE
where $a$~is a constant.  Recall that all of~$\Go^2$, $\nuz$, and~$\nuT$ are
positive.  To the extent that $\ebar \ll \nuz\nuT$, the
bifurcation occurs essentially at the linear 
threshold.  We understand this to define the onset of the Dimits-shift
regime. For $\ebar \gg \nuz\nuT$, no zonostrophic instability occurs to the
left of the linear threshold.

For a quantitative calculation, we recall the theory of \GK\
noise.\cite{JAK_es,Nevins_noise}  The general theory of statistical
fluctuations in the presence of both particle discreteness and turbulence
is complicated\cite{Rose}; some discussion is given in \Ref{JAK_noise}.
There are at least two qualitative issues.  Most fundamentally, common
statistical 
closures such as the DIA\footnote{For many references to the DIA, see the
  review article (\Ref{JAK_PR}) and tutorial article (\Ref{JAK_tutorial}) by
  the second author.} are incorrect because they make an assumption about
Gaussian 
initial statistics that is incorrect in the presence of particle
discreteness\cite{Rose} (this can easily be seen from the form of the
thermal-equilibrium Gibbs distribution).  Also, the details of discrete
particle noise depend on the shape of the one-particle distribution
function~$f$.  Fluid descriptions of the kind pursued in the present
article assume that $f$~is a local Maxwellian.  That is not unreasonable
for the level of description to which we aspire here, particularly since
there is no turbulence in the regimes of either linear stability or the
Dimits shift, but it should be
revisited in the regime where the particle noise is strongly amplified
close to linear threshold.  That is left for future work.

Thus we shall proceed as follows.  

(i)~For $\k = 0$, we determine the equilibrium \GK\ noise level using the
noise calculations of 
Krommes\cite{JAK_es} and Nevins \etal,\cite{Nevins_noise} based on the
Rostoker superposition principle.  The latter calculations
are more appropriate for the present case since they were done using the
assumption of adiabatic electron response.

(ii)~We infer the forcing functions by balancing the forcing against the
Landau damping rates assumed in the basic ITG fluid mode, for given noise
level.  Thus, if a scalar random variable~$\psit$ obeys
\BE
\Total{\psit}{t} + \nu\psit = \ft(t),
\eq{psit_dot}
\EE
where $\ft(t)$ is Gaussian white noise with covariance $\<\f f(t)\f f(t')> =
2D\Dirac{t-t'}$, the \sstate\ solution for the second-order statistics of
the Langevin equation \EQ{psit_dot} is\footnote{When $\protect\psit$~is the
  random 
velocity of a Brownian test particle of mass~$M$ in a thermal bath of
temperature~$T$, \Eq{dpsit2} leads to the Einstein relation $T/M =
\Dv/\nu$, where $\Dv$~is the velocity-space diffusion coefficient.}
\BE
\<\f\psi^2> = D/\nu.
\eq{dpsit2}
\EE
Given~$\<\f\psit^2>$, $D$~is determined as $\nu\<\f\psit^2>$.  The
determination of the $n$-dimensional forcing matrix~$\mD$ is a simple
generalization of this result, as we shall explicate below.

(iii)~Finally, now knowing~$\mD$, we turn on~$\k$ and proceed as in the
earlier part of this 
article to calculate the noise-drive, $\k$-dependent fluctuation level and
then the onset of the zonostrophic instability.

\subsubsection{Constraints on the damping coefficients; the forcing matrix}
\label{constraints}

To carry out the above program, we first calculate the equilibrium
correlation matrix $\mC\equil \doteq \<\f\vpsi\,\f\vpsi\adjo>\equil$, where
$\vpsit \doteq (\zt,\,\Tt)\Tr$.  (We omit the dependence on~$\vk$ of this
and the other quantities in the following discussion.)
Because we work in the
electrostatic approximation, both components of~$\d\vpsi$ are driven by
the random potential~$\f\ph$.  The vorticity is linearly related to~$\f\ph$:
$\f\z = \cZ\f\ph$, where $\cZ \doteq 1 + \kperpbar^2$.  The fluctuating
temperature could in principle contain nonlinear contributions, but since
we restrict ourselves to weak coupling, the linear approximation is
adequate:  $\f T = \cT\f\ph$, where we shall calculate~$\cT$ from the
equilibrium \GKE\ below.  

The fact that the vector $\f\vpsi$ is linearly proportional to the
scalar~$\f\ph$ has important consequences for the properties of the
equilibrium correlation matrix.  Let $\f\vpsi = \vS\,\f\ph$, where $\vS$~is an
$n$-dimensional, non-zero scaling vector.  [For our specific problem, we
have $\vS = 
(\cZ,\,\cT)\Tr$.]  Thus $\mC\equil = \vS\,\vS\Tr\Cpp\equil$.  This matrix
has one 
positive 
eigenvalue, $\lo_+ = \norm{\vS}^2 \doteq \vS\adjo\.\vS$, with
associated eigenvector $\ve_+ = \vShat \doteq \vS/\norm{\vS}$.  The
remaining $n-1$ eigenvalues vanish.  This follows since $\vShat$~can be
taken to define the normal to an $(n-1)$-dimensional hyperplane; thus one
can find $n-1$ vectors~$\vqhat_i$ ($i = 2,\dots,n$) such that
$\vShat\adjo\.\vqhat_i = 0$.  It follows that $\mC\equil$~is diagonalized
by the unitary matrix
$\mU \doteq (\ve_+,\,\vqhat_2,\dots,\vqhat_n)$ according to
\BE
\Bar{\mC} \doteq \mU\adjo\.\mC\.\mU = \norm{\vS}^2
\begin{pmatrix}
1 & 0 & \dots & 0\\
0 & 0 & \dots & 0\\
\vdots & \vdots & \ddots & 0\\
0 & 0 & \dots & 0
\end{pmatrix}
\Cpp\equil.
\EE

We now want to work backwards and determine the associated forcing
matrix~$\mD$.   The equilibrium Langevin equation
\BE
	\Total{\f\vpsit}{t} + \mV\.\f\vpsit = \f\vf,
\EE
where $\mV$~contains the linear dissipation,
transforms to
\BE
	\Total{\f\vpsibar}{t} + \mVbar\.\f\vpsibar = \f\vfbar,
\eq{dvpsibar_dot}
\EE
where 
\BE
\f\vpsibar \doteq \norm{\vS}
\begin{pmatrix}
1\\0\\\vdots\\0
\end{pmatrix}\f\ph,
\quad
\f\vfbar \doteq \norm{\vS}
\begin{pmatrix}
1\\0\\\vdots\\0
\end{pmatrix}\f f_\ph
\EE
and $\mVbar
\doteq \mU\adjo\.\mV\.\mU$.  The fact that the time derivative and forcing
appear only in the first component of \Eq{dvpsibar_dot} places restrictions
on the form of the 
transformed dissipation matrix~$\mVbar$, and thus on the original~$\mV$.
Specifically, if $\mV$~is taken to be diagonal, as we did in writing
the model system \Eq{zT}, then one can show that the diagonal elements must
be equal.  To 
see this explicitly for $n = 2$, suppose that
\BE
	\mVbar = \begin{pmatrix}a & b\\c & d\end{pmatrix}.
\EE
The unitary transformation matrix (whose columns are the eigenvectors) can
be chosen to be
\BE
	\mU = \fr{1}{\D\ehalf}\begin{pmatrix}
\cZ & -\cT\conj\\
\cT & \cZ\conj
\end{pmatrix},
\EE
where $\D \doteq \abso{\cZ}^2 + \abso{\cT}^2 = \norm{\vS}^2$.  With
$\d\vpsibar = \mU\adjo\.\d\vpsi$, 
this leads to the transformed Langevin equations
\BALams
\Total{\f\ph}{t} + a\,\f\ph &= \f f_\ph,
\eq{L_a}
\\ 
0 + c\,\f\ph &= 0.
\EALams
Thus $c = 0$, with $b$ and~$d$ being arbitrary at this point.  Upon
transforming back, one finds
\BE
	\mV = \begin{pmatrix}
\abso{\cZ}^2a - \cZ\cT b + \abso{\cT}^2d
& \cZ\cT\conj(a - d) + \cZ^2 b\\
\cZ\conj\cT(a - d) - \cT^2 b
& \abso{\cT}^2a + \cZ\cT b + \abso{\cZ}^2 d
\end{pmatrix}.
\EE
The requirement that the off-diagonal elements vanish is easily seen to
imply $b = 0$ and $d = a$; thus $\mV = \mVbar = a\mone$.

Upon replacing~$a$ by the common damping rate~$\nu = \nuz = \nuT = \nubar$
and applying the result 
\EQ{dpsit2} to \Eq{L_a}, one determines
\BE
	\Dpp\equil = \nu\<\abso{\f\ph}^2>\equil.
\EE
The back transformation $\d\vf = \mU\.\d\vfbar$ leads to
\BE
	\mD\equil = \nu\mC\equil,
\eq{mD_eq}
\EE
where
\BE
	\mC\equil = \begin{pmatrix}
\abso{\cZ}^2 & \cZ\cT\conj\\
\cZ\conj\cT & \abso{\cT}^2
\end{pmatrix}
\Cpp\equil.
\EE

\subsubsection{The scaling coefficients}

To determine the scaling coefficient~$\cT$, we solve the \GKE\ linearized
(denoted by~$\D$) around thermal equilibrium,
\BE
	\Partial{\D F}{t} + \vpar\gradpar \D F +
	\fr{q}{m}\D\Epar\Partial{\FM}{\vpar} 
	= 0
\EE
to find
\BE
	\D F_i(\vkw) = Z\t\m1\fR{\kpar\vpar}{\w - \kpar\vpar +
	  i\e}\D\ph(\vkw)\FMi, 
\EE
where $Z$~is the atomic number and $\t \doteq \Ti/\Te$.
The fluctuating temperature is then
\BE
	\D\Ti = \Int d\vv\,\(\Half\mi v^2 - \Case32\Ti\)\D F_i.
\EE
Since we ignore FLR effects, the perpendicular part of $\half v^2 - \frac32
= (\half\vpar^2 -\half) + (\half\vperp^2 - 1)$ integrates away.  For the
parallel part, it is conventional in standard drift-wave theory to look for
modes with $\w \gg \kpar\vti$.  Upon expanding the denominator for small
$\kpar\vpar/\w$, one finds
\BE
	\cT(\vkw) \doteq \D\ph\m1(\vkw)\fR{\D\Ti(\vkw)}{\Ti} \approx
	Z\t^{-1}\fr{\kpar^2\vti^2}{\w^2} \ll 1.
\EE
Unfortunately, this expansion is not well justified for the ITG mode.
However, one can already see another issue, which is that any such~$\cT$,
calculated with or without expansion, will depend on~$\w$.  This goes
beyond the level of detail assumed in the above white-noise calculations,
which assume that the scaling coefficients depended only on~$\vk$.  In
problems for which the linear eigenfrequency is dominantly real, this
difficulty can be justifiably surmounted by replacing~$\w$ by the real mode
frequency~$\Wk$.  When $\w$~is purely imaginary, this procedure is less
justified, and to properly deal with the fact that $\abso{\w} \sim
\kpar\vti$, one should do a kinetic analysis.  Alternatively, qualitatively
correct results should obtain by ignoring~$\cT$ altogether, thus forcing
only the vorticity.

\subsubsection{The \GK\ noise level}

As a trivial modification of the work of Nevins \etal,\cite{Nevins_noise} one finds for the case of
ITG modes with adiabatic electrons~\cite{JAK_noise} the fluctuating noise
level 
\BE
\frac{\langle \delta \phi\, \delta \phi \rangle(\vk)}{8\pi} =
\fr{T_\mathrm{i0}/2}{\eGV(\vk)}
\fR{\kDi^2\Gamma(\kperp^2
  \ri^2)/\kD^2}{k^2(1 +
  \eGV\m1 \kDe^2/k^2)(1+k^2 \lD^2)}. 
\EE
Here $\kDs$ is the Debye length for species~$s$,
$\kD^2 \doteq \kDe^2 + \kDi^2$,  
$\Gamma(b) \doteq I_0(b) e^{-b}$, and the dielectric permittivity of the
gyrokinetic vacuum $\eGV$ is given by 
\BE
\eGV \approx 1 +
\fR{\kDi^2}{k^2}(1 - \Gamma). 
\EE 
In our normalized variables we have
\BAams
&\mathsf{C}^{(\mathrm{eq})}_{\varphi\varphi}(k_x, k_y) = \int \frac{d k_z}{2 \pi}
\fr{1}{\ni}
 \fR{\tau \dielperp}{ \eGV}
\NN\\
&\quad\times
\fR{\Gamma( \tau \kperp^2) }{\rho_*^2 k^2 (1 +
 \eGV\m1\dielperp/ k^2)(1+\tau+ \tau k^2/\dielperp)}
\EAams
with
\BE
\eGV \approx 1 + \left(\frac{\dielperp}{\tau k^2}\right)[1 - \Gamma(\t k^2)].
\EE 
Here, $\rho_* \doteq  \rs /a$, $\tau \doteq T_\mathrm{i0}/T_\mathrm{e0}$, $\dielperp \doteq \wpi^2 / \wci^2$, and $\ni$ denotes the ion density scaled by $\rs^3$. The zonostrophic instability problem is thus specified in terms of the 
 parameters~$\ni$,~$\rho_*$,~$\t$ and~$\dielperp$.
In the gyrokinetic regime\cite{JAK_es} $\dielperp \gg 1$, so in the cold-ion
limit one has $\lim_{\t \to 0}\eGV \approx \dielperp$.

\subsection{Solution of the neutral curve equation}
\label{Neutral}

A representative, numerically calculated neutral curve is displayed in
  \Fig{normal3}. 
It displays the expected qualitative features, including a
 critical zonal wave number~$q\crit$, a supercritical bifurcation, and a
 parabolic shape near onset.  Notice that
 the zonostrophic instability sets in for $q\crit\rs = \Order{1}$, not
 $q\crit/k  \ll 1$.

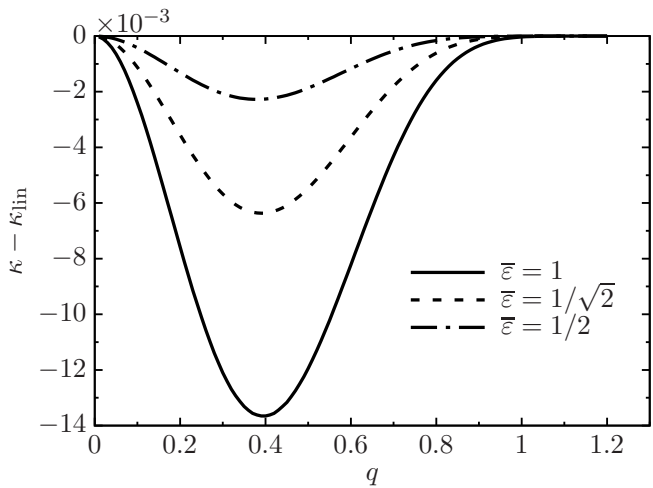
\FIGURE{normal3}{A a series of representative neutral curves for various  values of $\ebar$, calculated numerically for
 the \GK\ 
 noise spectrum given by Nevins \etal \cite{Nevins_noise} for $\t = 1$, $\rho_*=0.3$, $\ni=1$, $\dielperp=100$, $\cT=0$ and $\nu = |k_y|$ represents damping in a Landau-fluid closure. Curves are quantitatively similar for $\cT = -0.25$.}
 
Our principal purpose in displaying a numerically calculated neutral curve
is to demonstrate that our analytically derived dispersion relation is
robust and that no unexpected pathologies arise.  Clearly the entire model
is extremely crude, so our results cannot be quantitatively compared with
experiments or fully kinetic simulations.  

Missing from \Fig{normal3} is an inner, secondary stability curve for the
steady zonal solutions that emerge above 
the point of zonostrophic instability.  The existence of that boundary is
known from previous work, including most recently that of Parker and
Krommes,\cite{Parker-JAK_jets_PoP,Parker-JAK_NJP} who calculated it
numerically for the case of the modified \HME.  Calculation of 
that curve for the present model requires numerical work and is beyond the
scope of this paper.

\section{Discussion}

In summary, we have used the CE2 stochastic model to derive the dispersion
relation of the noise-driven 
zonostrophic instability for a simple two-field model of the
ion-temperature-gradient-driven mode, and we have numerically calculated
the neutral curve and found the first unstable zonal mode for a
representative noise spectrum.  

The principal goal of this work is to present a new interpretation of the
zonostrophic instability as being driven by discrete particle noise instead
of the more conventional interpretation as being due to coupling to
extrinsic turbulence.  While it is obvious that in realistic tokamak
microturbulence there is a plethora of modes in addition to the ITG mode,
coupling to those modes should not be necessary for a self-consistent
description of the behavior of the ITG mode itself as~$\k$ (the normalized
magnitude of the temperature gradient) is increased.  We have shown that
such a self-consistent description is possible when discrete particle noise
is included.  By introducing that noise, one is able to ``open up'' the
zonostrophic bifurcation that introduces the onset of the Dimits-shift
regime, which we have shown occurs just slightly below the linear
threshold.

Left undone is the extension of these results through the right-hand
boundary of the Dimits shift.  This requires addressing the secondary
stability boundary of the steady zonal flows that emerge from the
zonostrophic bifurcation.  If our interpretation is to be consistent with the known behavior observed in the simulations, that stability curve must
close off for sufficiently large~$\k$.  Such behavior --- the so-called Busse
balloon --- is known to occur for the closely analogous problem of
Rayleigh--Benard thermal convection.\cite{Busse67,Busse-Whitehead} Pursuing
this investigation would augment current understanding of the Dimits shift
and would provide a bridge between the sometimes arcane specialty of plasma
physics and a large and broad literature on bifurcation phenomena in
physical systems.

\acknowledgments

We would like to thank J.~Parker and G.~Hammett for many fruitful
discussions.  This work was supported by a NSERC PGS-D scholarship
and by U.S. DoE contract DE-AC02-09CH11466.

\appendix

\section{Realizability of the homogeneous solution}
\label{detC}

The CE2 closure deals only with first- and second-order statistics.
Assuming Gaussian forcing, the multivariate PDF of~$\pht$ and~$\Tt$ is a 2D
Gaussian.  Realizability of the \sstate\ solution for a nonsingular PDF
requires that 
\BE
\mC\up0 \doteq \begin{pmatrix}
\Cppo & \CpTo\\
\CpTo{}\conj & \CTTo
\end{pmatrix}
\EE
is positive definite\footnote{Strictly speaking, realizability only
  requires positive semidefiniteness.  Sylvester's criterion is then that
  all of the principal minor determinants must be non-negative.}
for all realizable forcings.  
For a matrix to be positive definite, Sylvester's criterion states that
its leading principal minor determinants must be positive.
Thus in the absence of any constraints on the forcing, one must satisfy
\BE
\Cpp > 0,
\quad
\Delta \doteq \Cppo\CTTo - \abso{\CpTo}^2 > 0.
\EE
If no mistakes have been made, realizability is guaranteed up to the linear
threshold because the covariance matrix has been derived from a set of
primitive amplitude equations driven by realizable random forcings; above
threshold, a homogeneous \Sstate\ does 
not exist.  As a partial check,
we consider the case with $\DpT'' = 0$.  Then from \Eq{mC0} one finds
that
\BE
\Cppo = (2\nubar\kperpbar^4\xihat)\m1
[(\xihat^2 - \nuT^2)\Dpp - \ky^2\ehat^2\DTT].
\EE
For $\xi < 0$ (linear stability), this is easily seen to be positive.

In terms of the real vector~$\vD \doteq (\Dpp,\, \DTT,\,\DpT')\Tr$,
the evaluation of~$\Delta$ for $\DpT'' = 0$ leads to the quadratic form
$F_3 = \vD\Tr\.\mS_3\.\vD$, where for this case
\BE
\mS_3 = \begin{pmatrix}
a & b & 0\\
b & d & 0\\
0 & 0 & f
\end{pmatrix};
\EE
thus $F_3 = a\Dpp^2 + 2b\Dpp\DTT + d\,\DTT^2 + f\DpT^{\prime2}$ and the
coefficients~$a$, $b$, $d$, and~$f$ can be obtained from \Eqs{CpT0'} and
\EQ{mC0}.  After some algebra, one finds
\BALams
a &= C(\nubar\abso{\xi})\m1(-\ky^2\k^2\xi),
\\
b &= C(\nubar\abso{\xi})\m1\kbar^4\[\xi - \half(\nuz^2 + \nuT^2)\]\xi,
\\
d &= C(\nubar\abso{\xi})\m1\kbar^4(-\ky^2\e^2\xi),
\\
f &= -(\nubar\kbar^2)\m1,
\EALams
where $C \doteq (2\kbar^4)\m1$.
In the region of linear stability ($\xi < 0$), we observe that $a$, $b$,
and~$d$ are positive, while $f < 0$.  
The criteria that $F_3$ be positive definite are
\BE
a > 0,
\quad
\D_2 \doteq ad-b^2 > 0,
\quad
\D_3 \doteq \D_2 f > 0.
\EE
 After more algebra, one finds
\BE
\D_2 = C^2[\xihat - (\nuz - \nuT)^2] < 0,
\EE
and then $\D_3 > 0$.  Thus the submatrix related to~$\Dpp$ and~$\DTT$
violates unconstrained positive definiteness.  However, realizabiity
constrains~$\Dpp$ and~$\DTT$ to be positive.  Since all of~$a$, $b$ and~$d$
are positive, the subform~$F_2$ is positive.  We have thus verified that
$\mC$~is realizable for realizable forcing (for the special case $\DpT'' =
0$).

To interpret the fact that $\D_2 < 0$, note that 
 the eigenvalues of~$\mS_2$
are $\lo = \Sigma \pm (\Sigma^2 - 4\D_2)\ehalf$, where $\Sigma \doteq a+d >
0$.  Thus one eigenvalue is negative.  The associated eigenvector satisfies
 $\DTT = -b\m1(a-\lo_-)\Dpp < 0$.  Such an unrealizable forcing 
would violate the condition that the cross-correlation coefficient
 between~$\f\ph$ and~$\f T$ must be less than~1 in absolute value.

\section{Details of the dispersion relation}
\label{Dispersion}

For all $\nu$'s equal, 
the matrix~$\mMhat$ and source vectors introduced in \Eq{MvC} have the form
\begin{widetext}
\BE
\mMhat \doteq 
\begin{pmatrix}
-(\l + 2)\hp^2\hm^2 & -i\ky\ehat\hp^2 & i\ky \ehat\hm^2 & 0\\
-i\ky\khat\hp^2 & (\l + 2)\hp^2 & 0 & -i\ky\ehat\\
i\ky\khat\hm^2 & 0 & (\l +2\nubar)\hm^2 & i\ky\ehat\\
0 & -i\ky\khat & i\ky\khat & -(\l + 2)
\end{pmatrix},
\EE
\BE
\vshat_U \doteq \fr{i\ky}{\nu}
\begin{pmatrix}
\hp^2(\hp^2 - q^2)\Cppp - \hm^2(\hm^2 - q^2)\Cppm\\
-\hp^2\CpTp + (\hm^2 - q^2)\CpTm\\
-[-\hm^2\CTpm + (\hp^2 - q^2)\CTpp]\\
\CTTp - \CTTm
\end{pmatrix},
\quad
\vshat_T \doteq \fr{\ky q}{\nu}
\begin{pmatrix}
0\\
-\hp^2\Cppp\\
\hm^2\Cppm\\
-(\CpTm - \CTpp)
\end{pmatrix}.
\EE

For the formalism to make sense, $\mMhat$~must be invertible, \ie, its
determinant must not vanish.  One finds
\BE
\D \doteq \det{\mMhat} = \hp^4\hm^4(\l + 2)^4
- 2\hp^2\hm^2(\hp^2 + \hm^2)(\xihat + 1)(\l + 2)^2
+ (\hp^2 - \hm^2)^2(\xihat + 1)^2.
\EE
\end{widetext}
There is no requirement that~$\D$ be positive definite; one needs only that
it not vanish at the point of zonostrophic bifurcation $\l = 0$.  This is not
expected since $\xis$~depends on the zonal-flow damping rates, which do not
appear in~$\mDhat$.  Some general properties of~$\D(\xihat,q) \equiv
\D(\l=0,\xihat,q)$ are easy to 
determine.  It can be shown to depend only on~$q^2$.  For
$\xihat = -1$, one has 
\BE
	\D(-1,q) = 16\hp^4\hm^4 > 0.
\EE
$\D$~has a minimum \wrt~$\xihat$ at
\BE
\xihat = \fr{4\hp^2 +\hm^2(\hp^2 + \hm^2)}{\kperpbar^2(\hp^2 - \hm^2)^4} > 0.
\EE
Its derivative at the linear threshold $\xihat = 0$ is
\BE
	\Partial{\D(\xihat,q)}{\xihat}\At{\xihat = 0} = -i\kperpbar^2\hp^2\hm^2(\hp^2 +
	\hm^2) < 0 
\EE
and its second derivative is
\BE
	\Partial{^2\D}{\xihat^2} = 2\kperpbar^4(\hp^2 - \hm^2)^2 > 0.
\EE
$\D(0,q)$~vanishes for $q = 0$ ($\hp^2 = \hm^2 = \kperpbar^2$), and its
derivative \wrt~$q^2$ is
\BE
	\Partial{\D(0,q)}{q^2} = -12\kx^2\kperpbar^4 < 0.
\EE
Thus except for $q = 0$ the determinant is negative at the linear
threshold.  Given that the second $\xihat$~derivative is uniformly
positive, one concludes that $\D$~must change sign somewhere in the stable
region.  

The dispersion relation can be simplified by using the transformation $C_{ij}^{(\pm)}(k_x, k_y) = C_{ji}^{(\mp)}(-k_x,-k_y)$. Once all equilibria are expressed at a single point $(k_x - q/2, k_y)$, the transformation $k_x \rightarrow k_x + q /2$ is then performed. After defining 
\begin{align*}
  	\Bar{h} 	&\doteq \kperpbar^2 + 2 k_x  q + q^2,
\\ 	\Omega  	&\doteq k_y^2 \epsilon \kappa(\Bar{k}^2+ \Bar{h}^2) - (\lambda + 2 \nu)^2 \Bar{k}^2\Bar{h}^2,
\\	\Gamma_{k} 	&\doteq \Omega - 2 k_y^2 \kappa \epsilon  \Bar{k}^2, 
\\	\Gamma_{h} 	&\doteq \Omega - 2 k_y^2 \kappa \epsilon  \Bar{h}^2, 
 \end{align*}
one finds that the determinant can be written as
\BE
 \Delta' \doteq \Omega^2 - 4 k_y^2 \epsilon^2 \kappa^2 \Bar{k}^2\Bar{h}^2
\EE 
and the dispersion relation becomes
\BE
(\lambda + \nu_\zeta^\mathrm{Z} - iq B)(\lambda +
    \nu_T^\mathrm{Z} - q D) - iq^2 AC = 0,   
\EE
where
\begin{widetext}
\begin{align*}
	A &\doteq  i q \epsilon \int \fr{d k_x\, d k_y}{(2\pi)^2} \, (k_x
	+q/2)k_y^3 
	\overline{k}^2 \Delta'[\Gamma_h   C_{\varphi\varphi}^{(0)}  +
	  2 i \epsilon   k_y \overline{h}^2(\lambda + 2 \nu) C_{\varphi
	    T}^{(0)}  ], 
\\	B &\doteq 2i\int \fr{d k_x\, d k_y}{(2\pi)^2} \,(k_x + q/2) k_y^2
	\Delta' [ 
  \Omega(\lambda+2\nu) \overline{k}^2
  (\overline{k}^2-q^2)C_{\varphi\varphi}^{(0)}  -2k_y^2\epsilon^2 (\lambda
  +2\nu)\overline{k}^2\overline{h}^2 C_{TT}^{(0)}  
 \\&\qquad \qquad -i k_y \epsilon\Gamma_k(\overline{k}^2 - q^2)C_{\varphi
	T}^{(0)} + i k_y \epsilon \overline{k}^2\Gamma_h C_{T
	\varphi}^{(0)} ], 
\\	C &\doteq 2 \int \fr{d k_x\, d k_y}{(2\pi)^2}\, k_y^2 \Delta' [-2
	\kappa(\lambda+2\nu)^2 
	k_y    \overline{k}^4\overline{h}^2
	(\overline{k}^2-q^2)C_{\varphi\varphi}^{(0)}  + k_y
	\epsilon(\Bar{h}^2\Gamma_k + \Bar{k}^2\Gamma_h ) C_{TT}^{(0)} 
\\ &  \qquad \qquad +   i(\lambda + 2   \nu)\overline{k}^2\overline{h}^2
	\Gamma_k C_{T \varphi}^{(0)}   -   i (\lambda + 2
	\nu)\overline{k}^2(\overline{k}^2 - q^2)\Gamma_h C_{\varphi
	T}^{(0)}  ], 
\\	D &\doteq q\int \fr{d k_x\, d k_y}{(2\pi)^2}\, k_y^2\Delta'  [
	(\lambda + 
	2\nu) 
	\overline{k}^2 \overline{h}^2  \Gamma_k C_{\varphi\varphi}^{(0)} -i
	k_y \epsilon (\overline{h}^2\Gamma_k  + \overline{k}^2\Gamma_h )
	C_{\varphi T}^{(0)} ]. 
\end{align*}
\end{widetext}
For  $\mathcal{Z}\mathcal{T}^*= \pm \mathcal{Z}^*\mathcal{T}$ and
$D_{ij}(k_x,k_y) = D_{ji}(k_x,-k_y)$, the dispersion relation is
real-valued. If either $\mathcal{Z}$ or  $\mathcal{T}$ are set to zero,
then  $A = C = 0$ identically. In order for the dispersion relation to be
satisfied, one must then solve 
\BE
(\lambda + \nu_\zeta^\mathrm{Z} - iq B)(\lambda +
  \nu_T^\mathrm{Z} - q D) = 0. 
\EE
One can recover the dispersion relation found by
Parker~\cite{Parker-JAK_NJP} in the flat-density limit ($\beta = 0$ in that
reference) by  forcing only the vorticity ($\mathcal{T}$ = 0), turning off
the linear coupling terms ($\kappa = \epsilon = 0$), and solving for the
first branch  $\lambda + \nu_\zeta^\mathrm{Z} - iq B = 0$. 

\def\pages#1#2{#1}
\def\range#1{#1 aggaa \}}
\def\rangeo#1-#2{#1--#2 aaa\}\}\}}
\def\press{in press}

\bibliography{/u/krommes/Tex/jak}

\begin{thebibliography}{43}%
\makeatletter
\providecommand \@ifxundefined [1]{%
 \@ifx{#1\undefined}
}%
\providecommand \@ifnum [1]{%
 \ifnum #1\expandafter \@firstoftwo
 \else \expandafter \@secondoftwo
 \fi
}%
\providecommand \@ifx [1]{%
 \ifx #1\expandafter \@firstoftwo
 \else \expandafter \@secondoftwo
 \fi
}%
\providecommand \natexlab [1]{#1}%
\providecommand \enquote  [1]{``#1''}%
\providecommand \bibnamefont  [1]{#1}%
\providecommand \bibfnamefont [1]{#1}%
\providecommand \citenamefont [1]{#1}%
\providecommand \href@noop [0]{\@secondoftwo}%
\providecommand \href [0]{\begingroup \@sanitize@url \@href}%
\providecommand \@href[1]{\@@startlink{#1}\@@href}%
\providecommand \@@href[1]{\endgroup#1\@@endlink}%
\providecommand \@sanitize@url [0]{\catcode `\\12\catcode `\$12\catcode
  `\&12\catcode `\#12\catcode `\^12\catcode `\_12\catcode `\%12\relax}%
\providecommand \@@startlink[1]{}%
\providecommand \@@endlink[0]{}%
\providecommand \url  [0]{\begingroup\@sanitize@url \@url }%
\providecommand \@url [1]{\endgroup\@href {#1}{\urlprefix }}%
\providecommand \urlprefix  [0]{URL }%
\providecommand \Eprint [0]{\href }%
\providecommand \doibase [0]{http://dx.doi.org/}%
\providecommand \selectlanguage [0]{\@gobble}%
\providecommand \bibinfo  [0]{\@secondoftwo}%
\providecommand \bibfield  [0]{\@secondoftwo}%
\providecommand \translation [1]{[#1]}%
\providecommand \BibitemOpen [0]{}%
\providecommand \bibitemStop [0]{}%
\providecommand \bibitemNoStop [0]{.\EOS\space}%
\providecommand \EOS [0]{\spacefactor3000\relax}%
\providecommand \BibitemShut  [1]{\csname bibitem#1\endcsname}%
\let\auto@bib@innerbib\@empty
\bibitem [{\citenamefont {Srinivasan}\ and\ \citenamefont
  {Young}(2012)}]{Srinivasan-Young}%
  \BibitemOpen
  \bibfield  {author} {\bibinfo {author} {\bibfnamefont {K.}~\bibnamefont
  {Srinivasan}}\ and\ \bibinfo {author} {\bibfnamefont {W.~R.}\ \bibnamefont
  {Young}},\ }\href@noop {} {\bibfield  {journal} {\bibinfo  {journal}
  {J.~Atmos.\ Sci.}\ }\textbf {\bibinfo {volume} {69}},\ \bibinfo {pages}
  {\range{1633}} (\bibinfo {year} {2012})}\BibitemShut {NoStop}%
\bibitem [{\citenamefont {Farrell}\ and\ \citenamefont
  {Ioannou}(2015)}]{Farrell_Jets}%
  \BibitemOpen
  \bibfield  {author} {\bibinfo {author} {\bibfnamefont {B.~F.}\ \bibnamefont
  {Farrell}}\ and\ \bibinfo {author} {\bibfnamefont {P.~J.}\ \bibnamefont
  {Ioannou}},\ }in\ \href@noop {} {\emph {\bibinfo {booktitle} {Zonal Jets}}},\
  \bibinfo {editor} {edited by\ \bibinfo {editor} {\bibfnamefont
  {B.}~\bibnamefont {Galperin}}\ and\ \bibinfo {editor} {\bibfnamefont
  {P.}~\bibnamefont {Read}}}\ (\bibinfo  {publisher} {Cambridge University
  Press},\ \bibinfo {address} {Cambridge},\ \bibinfo {year} {2015})\ Chap.\
  \bibinfo {chapter} {V.2.2},\ \bibinfo {note} {\press}\BibitemShut {NoStop}%
\bibitem [{\citenamefont {Tobias}, \citenamefont {Dagon},\ and\ \citenamefont
  {Marston}(2011)}]{Tobias11}%
  \BibitemOpen
  \bibfield  {author} {\bibinfo {author} {\bibfnamefont {S.~M.}\ \bibnamefont
  {Tobias}}, \bibinfo {author} {\bibfnamefont {K.}~\bibnamefont {Dagon}}, \
  and\ \bibinfo {author} {\bibfnamefont {J.~B.}\ \bibnamefont {Marston}},\
  }\href@noop {} {\bibfield  {journal} {\bibinfo  {journal} {Astrophys.~J.}\
  }\textbf {\bibinfo {volume} {727}},\ \bibinfo {pages} {\range{127}} (\bibinfo
  {year} {2011})}\BibitemShut {NoStop}%
\bibitem [{\citenamefont {Tobias}\ and\ \citenamefont
  {Marston}(2013)}]{Tobias-Marston13}%
  \BibitemOpen
  \bibfield  {author} {\bibinfo {author} {\bibfnamefont {S.~M.}\ \bibnamefont
  {Tobias}}\ and\ \bibinfo {author} {\bibfnamefont {J.~B.}\ \bibnamefont
  {Marston}},\ }\href {\doibase
  \url{http://dx.doi.org/10.1103/PhysRevLett.110.104502}} {\bibfield  {journal}
  {\bibinfo  {journal} {Phys.\ Rev.\ Lett.}\ }\textbf {\bibinfo {volume}
  {110}},\ \bibinfo {pages} {\pages{104502}{5}} (\bibinfo {year}
  {2013})}\BibitemShut {NoStop}%
\bibitem [{\citenamefont {Marston}, \citenamefont {Qi},\ and\ \citenamefont
  {Tobias}(2016)}]{Marston_Jets}%
  \BibitemOpen
  \bibfield  {author} {\bibinfo {author} {\bibfnamefont {J.~B.}\ \bibnamefont
  {Marston}}, \bibinfo {author} {\bibfnamefont {W.}~\bibnamefont {Qi}}, \ and\
  \bibinfo {author} {\bibfnamefont {S.~M.}\ \bibnamefont {Tobias}},\ }in\
  \href@noop {} {\emph {\bibinfo {booktitle} {Zonal Jets}}},\ \bibinfo {editor}
  {edited by\ \bibinfo {editor} {\bibfnamefont {B.}~\bibnamefont {Galperin}}\
  and\ \bibinfo {editor} {\bibfnamefont {P.}~\bibnamefont {Read}}}\ (\bibinfo
  {publisher} {Cambridge University Press},\ \bibinfo {address} {Cambridge},\
  \bibinfo {year} {2016})\ Chap.\ \bibinfo {chapter} {V.1.2},\ \bibinfo {note}
  {\press}\BibitemShut {NoStop}%
\bibitem [{\citenamefont {Farrell}\ and\ \citenamefont
  {Ioannou}(2003)}]{Farrell03_SSST}%
  \BibitemOpen
  \bibfield  {author} {\bibinfo {author} {\bibfnamefont {B.~F.}\ \bibnamefont
  {Farrell}}\ and\ \bibinfo {author} {\bibfnamefont {P.~J.}\ \bibnamefont
  {Ioannou}},\ }\href@noop {} {\bibfield  {journal} {\bibinfo  {journal}
  {J.~Atmos.\ Sci.}\ }\textbf {\bibinfo {volume} {60}},\ \bibinfo {pages}
  {\range{2101}} (\bibinfo {year} {2003})}\BibitemShut {NoStop}%
\bibitem [{\citenamefont {Farrell}\ and\ \citenamefont
  {Ioannou}(2009)}]{Farrell_Dw_Zf}%
  \BibitemOpen
  \bibfield  {author} {\bibinfo {author} {\bibfnamefont {B.~F.}\ \bibnamefont
  {Farrell}}\ and\ \bibinfo {author} {\bibfnamefont {P.~J.}\ \bibnamefont
  {Ioannou}},\ }\href@noop {} {\bibfield  {journal} {\bibinfo  {journal}
  {Phys.\ Plasmas}\ }\textbf {\bibinfo {volume} {16}},\ \bibinfo {pages}
  {\pages{112903}{19}} (\bibinfo {year} {2009})}\BibitemShut {NoStop}%
\bibitem [{\citenamefont {Krommes}(2015)}]{JAK_tutorial}%
  \BibitemOpen
  \bibfield  {author} {\bibinfo {author} {\bibfnamefont {J.~A.}\ \bibnamefont
  {Krommes}},\ }\href@noop {} {\bibfield  {journal} {\bibinfo  {journal}
  {J.~Plasma Phys.}\ }\textbf {\bibinfo {volume} {81}},\ \bibinfo {pages}
  {\range{1}} (\bibinfo {year} {2015})}\BibitemShut {NoStop}%
\bibitem [{\citenamefont {Kadomtsev}(1965)}]{Kadomtsev}%
  \BibitemOpen
  \bibfield  {author} {\bibinfo {author} {\bibfnamefont {B.~B.}\ \bibnamefont
  {Kadomtsev}},\ }\href@noop {} {\emph {\bibinfo {title} {Plasma Turbulence}}}\
  (\bibinfo  {publisher} {Academic},\ \bibinfo {address} {New York},\ \bibinfo
  {year} {1965})\ \bibinfo {note} {translated by L.~C. Ronson from
  \emph{Problems in Plasma Theory}, Vol.~4, edited by M.~A. Leontovich
  (Atomizdat, Moscow, 1964) pp.~188--339. Translation edited by M.~C.
  Rusbridge}\BibitemShut {NoStop}%
\bibitem [{\citenamefont {Krommes}(2002)}]{JAK_PR}%
  \BibitemOpen
  \bibfield  {author} {\bibinfo {author} {\bibfnamefont {J.~A.}\ \bibnamefont
  {Krommes}},\ }\href {\doibase
  \url{http://dx.doi.org/10.1016/S0370-1573(01)00066-7}} {\bibfield  {journal}
  {\bibinfo  {journal} {Phys.\ Rep.}\ }\textbf {\bibinfo {volume} {360}},\
  \bibinfo {pages} {\range{1}} (\bibinfo {year} {2002})}\BibitemShut {NoStop}%
\bibitem [{\citenamefont {Dimits}\ \emph {et~al.}(2000)\citenamefont {Dimits},
  \citenamefont {Bateman}, \citenamefont {Beer}, \citenamefont {Cohen},
  \citenamefont {Dorland}, \citenamefont {Hammett}, \citenamefont {Kim},
  \citenamefont {Kinsey}, \citenamefont {Kotschenreuther}, \citenamefont
  {Kritz}, \citenamefont {Lao}, \citenamefont {Mandrekas}, \citenamefont
  {Nevins}, \citenamefont {Parker}, \citenamefont {Redd}, \citenamefont
  {Shumaker}, \citenamefont {Sydora},\ and\ \citenamefont
  {Weiland}}]{Dimits_shift}%
  \BibitemOpen
  \bibfield  {author} {\bibinfo {author} {\bibfnamefont {A.~M.}\ \bibnamefont
  {Dimits}}, \bibinfo {author} {\bibfnamefont {G.}~\bibnamefont {Bateman}},
  \bibinfo {author} {\bibfnamefont {M.~A.}\ \bibnamefont {Beer}}, \bibinfo
  {author} {\bibfnamefont {B.~I.}\ \bibnamefont {Cohen}}, \bibinfo {author}
  {\bibfnamefont {W.}~\bibnamefont {Dorland}}, \bibinfo {author} {\bibfnamefont
  {G.~W.}\ \bibnamefont {Hammett}}, \bibinfo {author} {\bibfnamefont
  {C.}~\bibnamefont {Kim}}, \bibinfo {author} {\bibfnamefont {J.~E.}\
  \bibnamefont {Kinsey}}, \bibinfo {author} {\bibfnamefont {M.}~\bibnamefont
  {Kotschenreuther}}, \bibinfo {author} {\bibfnamefont {A.~H.}\ \bibnamefont
  {Kritz}}, \bibinfo {author} {\bibfnamefont {L.~L.}\ \bibnamefont {Lao}},
  \bibinfo {author} {\bibfnamefont {J.}~\bibnamefont {Mandrekas}}, \bibinfo
  {author} {\bibfnamefont {W.~M.}\ \bibnamefont {Nevins}}, \bibinfo {author}
  {\bibfnamefont {S.~E.}\ \bibnamefont {Parker}}, \bibinfo {author}
  {\bibfnamefont {A.~J.}\ \bibnamefont {Redd}}, \bibinfo {author}
  {\bibfnamefont {D.~E.}\ \bibnamefont {Shumaker}}, \bibinfo {author}
  {\bibfnamefont {R.}~\bibnamefont {Sydora}}, \ and\ \bibinfo {author}
  {\bibfnamefont {J.}~\bibnamefont {Weiland}},\ }\href@noop {} {\bibfield
  {journal} {\bibinfo  {journal} {Phys.\ Plasmas}\ }\textbf {\bibinfo {volume}
  {7}},\ \bibinfo {pages} {\range{969}} (\bibinfo {year} {2000})}\BibitemShut
  {NoStop}%
\bibitem [{\citenamefont {Rogers}, \citenamefont {Dorland},\ and\ \citenamefont
  {Kotschenreuther}(2000)}]{Rogers00}%
  \BibitemOpen
  \bibfield  {author} {\bibinfo {author} {\bibfnamefont {B.~N.}\ \bibnamefont
  {Rogers}}, \bibinfo {author} {\bibfnamefont {W.}~\bibnamefont {Dorland}}, \
  and\ \bibinfo {author} {\bibfnamefont {M.}~\bibnamefont {Kotschenreuther}},\
  }\href@noop {} {\bibfield  {journal} {\bibinfo  {journal} {Phys.\ Rev.\
  Lett.}\ }\textbf {\bibinfo {volume} {85}},\ \bibinfo {pages} {\range{5336}}
  (\bibinfo {year} {2000})}\BibitemShut {NoStop}%
\bibitem [{\citenamefont {Kolesnikov}\ and\ \citenamefont
  {Krommes}(2005{\natexlab{a}})}]{Kolesnikov_PRL}%
  \BibitemOpen
  \bibfield  {author} {\bibinfo {author} {\bibfnamefont {R.~A.}\ \bibnamefont
  {Kolesnikov}}\ and\ \bibinfo {author} {\bibfnamefont {J.~A.}\ \bibnamefont
  {Krommes}},\ }\href@noop {} {\bibfield  {journal} {\bibinfo  {journal}
  {Phys.\ Rev.\ Lett.}\ }\textbf {\bibinfo {volume} {94}},\ \bibinfo {pages}
  {\pages{235002}{4}} (\bibinfo {year} {2005}{\natexlab{a}})}\BibitemShut
  {NoStop}%
\bibitem [{\citenamefont {Kolesnikov}\ and\ \citenamefont
  {Krommes}(2005{\natexlab{b}})}]{Kolesnikov_PoP}%
  \BibitemOpen
  \bibfield  {author} {\bibinfo {author} {\bibfnamefont {R.~A.}\ \bibnamefont
  {Kolesnikov}}\ and\ \bibinfo {author} {\bibfnamefont {J.~A.}\ \bibnamefont
  {Krommes}},\ }\href@noop {} {\bibfield  {journal} {\bibinfo  {journal}
  {Phys.\ Plasmas}\ }\textbf {\bibinfo {volume} {12}},\ \bibinfo {pages}
  {\pages{122302}{25}} (\bibinfo {year} {2005}{\natexlab{b}})}\BibitemShut
  {NoStop}%
\bibitem [{Note1()}]{Note1}%
  \BibitemOpen
  \bibinfo {note} {An introductory review article on gyrokinetics that contains
  references to more specialized reviews and original papers is by J.~A.
  Krommes, Annu.\ Ref.\ Fluid Mech.\protect \textbf {44}, 175
  (2012).}\BibitemShut {Stop}%
\bibitem [{\citenamefont {Krommes}, \citenamefont {Lee},\ and\ \citenamefont
  {Oberman}(1986)}]{JAK_es}%
  \BibitemOpen
  \bibfield  {author} {\bibinfo {author} {\bibfnamefont {J.~A.}\ \bibnamefont
  {Krommes}}, \bibinfo {author} {\bibfnamefont {W.~W.}\ \bibnamefont {Lee}}, \
  and\ \bibinfo {author} {\bibfnamefont {C.}~\bibnamefont {Oberman}},\ }\href
  {\url{http://dx.doi.org/10.1063/1.865534}} {\bibfield  {journal} {\bibinfo
  {journal} {Phys.\ Fluids}\ }\textbf {\bibinfo {volume} {29}},\ \bibinfo
  {pages} {\range{2421}} (\bibinfo {year} {1986})}\BibitemShut {NoStop}%
\bibitem [{\citenamefont {Nevins}\ \emph {et~al.}(2005)\citenamefont {Nevins},
  \citenamefont {Hammett}, \citenamefont {Dimits}, \citenamefont {Dorland},\
  and\ \citenamefont {Shumaker}}]{Nevins_noise}%
  \BibitemOpen
  \bibfield  {author} {\bibinfo {author} {\bibfnamefont {W.~M.}\ \bibnamefont
  {Nevins}}, \bibinfo {author} {\bibfnamefont {G.~W.}\ \bibnamefont {Hammett}},
  \bibinfo {author} {\bibfnamefont {A.~M.}\ \bibnamefont {Dimits}}, \bibinfo
  {author} {\bibfnamefont {W.}~\bibnamefont {Dorland}}, \ and\ \bibinfo
  {author} {\bibfnamefont {D.~E.}\ \bibnamefont {Shumaker}},\ }\href@noop {}
  {\bibfield  {journal} {\bibinfo  {journal} {Phys.\ Plasmas}\ }\textbf
  {\bibinfo {volume} {12}},\ \bibinfo {pages} {\pages{122305}{16}} (\bibinfo
  {year} {2005})}\BibitemShut {NoStop}%
\bibitem [{\citenamefont {Parker}\ and\ \citenamefont
  {Krommes}(2013)}]{Parker-JAK_jets_PoP}%
  \BibitemOpen
  \bibfield  {author} {\bibinfo {author} {\bibfnamefont {J.~B.}\ \bibnamefont
  {Parker}}\ and\ \bibinfo {author} {\bibfnamefont {J.~A.}\ \bibnamefont
  {Krommes}},\ }\href@noop {} {\bibfield  {journal} {\bibinfo  {journal}
  {Phys.\ Plasmas}\ }\textbf {\bibinfo {volume} {20}},\ \bibinfo {pages}
  {\pages{100703}{4}} (\bibinfo {year} {2013})}\BibitemShut {NoStop}%
\bibitem [{\citenamefont {Parker}\ and\ \citenamefont
  {Krommes}(2014)}]{Parker-JAK_NJP}%
  \BibitemOpen
  \bibfield  {author} {\bibinfo {author} {\bibfnamefont {J.~B.}\ \bibnamefont
  {Parker}}\ and\ \bibinfo {author} {\bibfnamefont {J.~A.}\ \bibnamefont
  {Krommes}},\ }\href {\doibase \url{dx.doi.org/10.1088/1367-2630/16/3/035006}}
  {\bibfield  {journal} {\bibinfo  {journal} {New J. Phys.}\ }\textbf {\bibinfo
  {volume} {16}},\ \bibinfo {pages} {\pages{035006}{28}} (\bibinfo {year}
  {2014})}\BibitemShut {NoStop}%
\bibitem [{\citenamefont {Parker}\ and\ \citenamefont
  {Krommes}(2016)}]{Parker-JAK_Jets}%
  \BibitemOpen
  \bibfield  {author} {\bibinfo {author} {\bibfnamefont {J.~B.}\ \bibnamefont
  {Parker}}\ and\ \bibinfo {author} {\bibfnamefont {J.~A.}\ \bibnamefont
  {Krommes}},\ }in\ \href@noop {} {\emph {\bibinfo {booktitle} {Zonal Jets}}},\
  \bibinfo {editor} {edited by\ \bibinfo {editor} {\bibfnamefont
  {B.}~\bibnamefont {Galperin}}\ and\ \bibinfo {editor} {\bibfnamefont
  {P.}~\bibnamefont {Read}}}\ (\bibinfo  {publisher} {Cambridge University
  Press},\ \bibinfo {address} {Cambridge},\ \bibinfo {year} {2016})\ Chap.\
  \bibinfo {chapter} {V.2.4},\ \bibinfo {note} {\press}\BibitemShut {NoStop}%
\bibitem [{\citenamefont {Chen}\ and\ \citenamefont
  {Cheng}(1980)}]{Chen-Cheng}%
  \BibitemOpen
  \bibfield  {author} {\bibinfo {author} {\bibfnamefont {L.}~\bibnamefont
  {Chen}}\ and\ \bibinfo {author} {\bibfnamefont {C.~Z.}\ \bibnamefont
  {Cheng}},\ }\href@noop {} {\bibfield  {journal} {\bibinfo  {journal} {Phys.\
  Fluids}\ }\textbf {\bibinfo {volume} {23}},\ \bibinfo {pages} {\range{2242}}
  (\bibinfo {year} {1980})}\BibitemShut {NoStop}%
\bibitem [{\citenamefont {Ottaviani}\ \emph {et~al.}(1990)\citenamefont
  {Ottaviani}, \citenamefont {Romanelli}, \citenamefont {Benzi}, \citenamefont
  {Briscolini}, \citenamefont {Santangelo},\ and\ \citenamefont
  {Succi}}]{Ottaviani_2field}%
  \BibitemOpen
  \bibfield  {author} {\bibinfo {author} {\bibfnamefont {M.}~\bibnamefont
  {Ottaviani}}, \bibinfo {author} {\bibfnamefont {F.}~\bibnamefont
  {Romanelli}}, \bibinfo {author} {\bibfnamefont {R.}~\bibnamefont {Benzi}},
  \bibinfo {author} {\bibfnamefont {M.}~\bibnamefont {Briscolini}}, \bibinfo
  {author} {\bibfnamefont {P.}~\bibnamefont {Santangelo}}, \ and\ \bibinfo
  {author} {\bibfnamefont {S.}~\bibnamefont {Succi}},\ }\href@noop {}
  {\bibfield  {journal} {\bibinfo  {journal} {Phys.\ Fluids~B}\ }\textbf
  {\bibinfo {volume} {2}},\ \bibinfo {pages} {67} (\bibinfo {year}
  {1990})}\BibitemShut {NoStop}%
\bibitem [{\citenamefont {Hammett}\ \emph {et~al.}(1993)\citenamefont
  {Hammett}, \citenamefont {Beer}, \citenamefont {Dorland}, \citenamefont
  {Cowley},\ and\ \citenamefont {Smith}}]{Hammett_developments}%
  \BibitemOpen
  \bibfield  {author} {\bibinfo {author} {\bibfnamefont {G.~W.}\ \bibnamefont
  {Hammett}}, \bibinfo {author} {\bibfnamefont {M.~A.}\ \bibnamefont {Beer}},
  \bibinfo {author} {\bibfnamefont {W.}~\bibnamefont {Dorland}}, \bibinfo
  {author} {\bibfnamefont {S.~C.}\ \bibnamefont {Cowley}}, \ and\ \bibinfo
  {author} {\bibfnamefont {S.~A.}\ \bibnamefont {Smith}},\ }\href@noop {}
  {\bibfield  {journal} {\bibinfo  {journal} {Plasma Phys.\ Control.\ Fusion}\
  }\textbf {\bibinfo {volume} {35}},\ \bibinfo {pages} {\range{973}} (\bibinfo
  {year} {1993})}\BibitemShut {NoStop}%
\bibitem [{\citenamefont {Candy}\ and\ \citenamefont {Waltz}(2003)}]{GYRO_PRL}%
  \BibitemOpen
  \bibfield  {author} {\bibinfo {author} {\bibfnamefont {J.}~\bibnamefont
  {Candy}}\ and\ \bibinfo {author} {\bibfnamefont {R.~E.}\ \bibnamefont
  {Waltz}},\ }\href@noop {} {\bibfield  {journal} {\bibinfo  {journal} {Phys.\
  Rev.\ Lett.}\ }\textbf {\bibinfo {volume} {91}},\ \bibinfo {pages} {045001 (4
  pages)} (\bibinfo {year} {2003})}\BibitemShut {NoStop}%
\bibitem [{\citenamefont {Biglari}, \citenamefont {Diamond},\ and\
  \citenamefont {Rosenbluth}(1989)}]{Biglari_toroidal_ITG}%
  \BibitemOpen
  \bibfield  {author} {\bibinfo {author} {\bibfnamefont {H.}~\bibnamefont
  {Biglari}}, \bibinfo {author} {\bibfnamefont {P.~H.}\ \bibnamefont
  {Diamond}}, \ and\ \bibinfo {author} {\bibfnamefont {M.~N.}\ \bibnamefont
  {Rosenbluth}},\ }\href@noop {} {\bibfield  {journal} {\bibinfo  {journal}
  {Phys.\ Fluids~B}\ }\textbf {\bibinfo {volume} {1}},\ \bibinfo {pages} {109}
  (\bibinfo {year} {1989})}\BibitemShut {NoStop}%
\bibitem [{\citenamefont {Romanelli}\ and\ \citenamefont
  {Briguglio}(1990)}]{Romanelli_Briguglio}%
  \BibitemOpen
  \bibfield  {author} {\bibinfo {author} {\bibfnamefont {F.}~\bibnamefont
  {Romanelli}}\ and\ \bibinfo {author} {\bibfnamefont {S.}~\bibnamefont
  {Briguglio}},\ }\href@noop {} {\bibfield  {journal} {\bibinfo  {journal}
  {Phys.\ Fluids~B}\ }\textbf {\bibinfo {volume} {2}},\ \bibinfo {pages} {754}
  (\bibinfo {year} {1990})}\BibitemShut {NoStop}%
\bibitem [{\citenamefont {Ottaviani}\ \emph {et~al.}(1997)\citenamefont
  {Ottaviani}, \citenamefont {Beer}, \citenamefont {Cowley}, \citenamefont
  {Horton},\ and\ \citenamefont {Krommes}}]{Ottaviani_ITP}%
  \BibitemOpen
  \bibfield  {author} {\bibinfo {author} {\bibfnamefont {M.}~\bibnamefont
  {Ottaviani}}, \bibinfo {author} {\bibfnamefont {M.}~\bibnamefont {Beer}},
  \bibinfo {author} {\bibfnamefont {S.}~\bibnamefont {Cowley}}, \bibinfo
  {author} {\bibfnamefont {W.}~\bibnamefont {Horton}}, \ and\ \bibinfo {author}
  {\bibfnamefont {J.}~\bibnamefont {Krommes}},\ }\href@noop {} {\bibfield
  {journal} {\bibinfo  {journal} {Phys.\ Rep.}\ }\textbf {\bibinfo {volume}
  {283}},\ \bibinfo {pages} {\range{121}} (\bibinfo {year} {1997})}\BibitemShut
  {NoStop}%
\bibitem [{\citenamefont {Krommes}\ and\ \citenamefont
  {Parker}(2015)}]{JAK-Parker_Jets}%
  \BibitemOpen
  \bibfield  {author} {\bibinfo {author} {\bibfnamefont {J.~A.}\ \bibnamefont
  {Krommes}}\ and\ \bibinfo {author} {\bibfnamefont {J.~B.}\ \bibnamefont
  {Parker}},\ }in\ \href@noop {} {\emph {\bibinfo {booktitle} {Zonal Jets}}},\
  \bibinfo {editor} {edited by\ \bibinfo {editor} {\bibfnamefont
  {B.}~\bibnamefont {Galperin}}\ and\ \bibinfo {editor} {\bibfnamefont
  {P.}~\bibnamefont {Read}}}\ (\bibinfo  {publisher} {Cambridge University
  Press},\ \bibinfo {address} {Cambridge},\ \bibinfo {year} {2015})\ Chap.\
  \bibinfo {chapter} {V.1.1},\ \bibinfo {note} {\press}\BibitemShut {NoStop}%
\bibitem [{\citenamefont {Kraichnan}(1959)}]{Kr59}%
  \BibitemOpen
  \bibfield  {author} {\bibinfo {author} {\bibfnamefont {R.~H.}\ \bibnamefont
  {Kraichnan}},\ }\href@noop {} {\bibfield  {journal} {\bibinfo  {journal}
  {J.~Fluid Mech.}\ }\textbf {\bibinfo {volume} {5}},\ \bibinfo {pages}
  {\range{497}} (\bibinfo {year} {1959})}\BibitemShut {NoStop}%
\bibitem [{\citenamefont {Leith}(1971)}]{Leith71}%
  \BibitemOpen
  \bibfield  {author} {\bibinfo {author} {\bibfnamefont {C.~E.}\ \bibnamefont
  {Leith}},\ }\href@noop {} {\bibfield  {journal} {\bibinfo  {journal}
  {J.~Atmos.\ Sci.}\ }\textbf {\bibinfo {volume} {28}},\ \bibinfo {pages}
  {\range{145}} (\bibinfo {year} {1971})}\BibitemShut {NoStop}%
\bibitem [{\citenamefont {Kraichnan}(1970)}]{Kr_convergents}%
  \BibitemOpen
  \bibfield  {author} {\bibinfo {author} {\bibfnamefont {R.~H.}\ \bibnamefont
  {Kraichnan}},\ }\href {\doibase
  \url{http://dx.doi.org/10.1017/S0022112070000587}} {\bibfield  {journal}
  {\bibinfo  {journal} {J.~Fluid Mech.}\ }\textbf {\bibinfo {volume} {41}},\
  \bibinfo {pages} {\range{189}} (\bibinfo {year} {1970})}\BibitemShut {NoStop}%
\bibitem [{\citenamefont {Kraichnan}(1971)}]{TFM}%
  \BibitemOpen
  \bibfield  {author} {\bibinfo {author} {\bibfnamefont {R.~H.}\ \bibnamefont
  {Kraichnan}},\ }\href {\doibase
  \url{http://dx.doi.org/10.1017/S0022112071001204}} {\bibfield  {journal}
  {\bibinfo  {journal} {J.~Fluid Mech.}\ }\textbf {\bibinfo {volume} {47}},\
  \bibinfo {pages} {\range{513}} (\bibinfo {year} {1971})}\BibitemShut {NoStop}%
\bibitem [{Note2()}]{Note2}%
  \BibitemOpen
  \bibinfo {note} {In this paper we use the standard plasma-physics slab
  coordinate system in which $x$~represents the direction of inhomogeneity (the
  radial coordinate in a tokamak) and $y$~represents the zonal direction (the
  poloidal direction in a tokamak). (In detail, in an actual tokamak the zonal
  flows are in the direction perpendicular to both~$x$ and~${\protect \bm
  {B}}$, \protect \textrm {i.e.}, mostly in the poloidal direction for large
  aspect ratio.) In geophysics, the roles of~$x$ and~$y$ are
  interchanged.}\BibitemShut {Stop}%
\bibitem [{\citenamefont {Ruiz}\ \emph {et~al.}()\citenamefont {Ruiz},
  \citenamefont {Parker}, \citenamefont {Shi},\ and\ \citenamefont
  {Dodin}}]{Ruiz-Parker}%
  \BibitemOpen
  \bibfield  {author} {\bibinfo {author} {\bibfnamefont {D.~E.}\ \bibnamefont
  {Ruiz}}, \bibinfo {author} {\bibfnamefont {J.~B.}\ \bibnamefont {Parker}},
  \bibinfo {author} {\bibfnamefont {E.~L.}\ \bibnamefont {Shi}}, \ and\
  \bibinfo {author} {\bibfnamefont {I.~Y.}\ \bibnamefont {Dodin}},\ }\href@noop
  {} {\enquote {\bibinfo {title} {Two corrections to the drift-wave kinetic
  equation in the context of zonal-flow physics},}\ }\bibinfo {note}
  {{s}ubmitted}\BibitemShut {NoStop}%
\bibitem [{\citenamefont {Cross}\ and\ \citenamefont
  {Greenside}(2009)}]{Cross-Greenside}%
  \BibitemOpen
  \bibfield  {author} {\bibinfo {author} {\bibfnamefont {M.}~\bibnamefont
  {Cross}}\ and\ \bibinfo {author} {\bibfnamefont {H.~S.}\ \bibnamefont
  {Greenside}},\ }\href@noop {} {\emph {\bibinfo {title} {Pattern Formation and
  Dynamics in Nonlinear Systems}}}\ (\bibinfo  {publisher} {Cambridge
  University Press},\ \bibinfo {address} {Cambridge},\ \bibinfo {year}
  {2009})\BibitemShut {NoStop}%
\bibitem [{\citenamefont {Rosenbluth}\ and\ \citenamefont
  {Hinton}(1998)}]{Rosenbluth-Hinton}%
  \BibitemOpen
  \bibfield  {author} {\bibinfo {author} {\bibfnamefont {M.~N.}\ \bibnamefont
  {Rosenbluth}}\ and\ \bibinfo {author} {\bibfnamefont {F.~L.}\ \bibnamefont
  {Hinton}},\ }\href@noop {} {\bibfield  {journal} {\bibinfo  {journal} {Phys.\
  Rev.\ Lett.}\ }\textbf {\bibinfo {volume} {80}},\ \bibinfo {pages}
  {\range{724}} (\bibinfo {year} {1998})}\BibitemShut {NoStop}%
\bibitem [{\citenamefont {Rose}(1979)}]{Rose}%
  \BibitemOpen
  \bibfield  {author} {\bibinfo {author} {\bibfnamefont {H.~A.}\ \bibnamefont
  {Rose}},\ }\href@noop {} {\bibfield  {journal} {\bibinfo  {journal}
  {J.~Stat.\ Phys.}\ }\textbf {\bibinfo {volume} {20}},\ \bibinfo {pages}
  {\range{415}} (\bibinfo {year} {1979})}\BibitemShut {NoStop}%
\bibitem [{\citenamefont {Krommes}(2007)}]{JAK_noise}%
  \BibitemOpen
  \bibfield  {author} {\bibinfo {author} {\bibfnamefont {J.~A.}\ \bibnamefont
  {Krommes}},\ }\href@noop {} {\bibfield  {journal} {\bibinfo  {journal}
  {Phys.\ Plasmas}\ }\textbf {\bibinfo {volume} {14}},\ \bibinfo {pages}
  {\pages{090501}{26}} (\bibinfo {year} {2007})}\BibitemShut {NoStop}%
\bibitem [{Note3()}]{Note3}%
  \BibitemOpen
  \bibinfo {note} {For many references to the DIA, see the review article
  (Ref.~\protect \rev@citealpnum {JAK_PR}) and tutorial article (Ref.~\protect
  \rev@citealpnum {JAK_tutorial}) by the second author.}\BibitemShut {Stop}%
\bibitem [{Note4()}]{Note4}%
  \BibitemOpen
  \bibinfo {note} {When $\protect \psit $~is the random velocity of a Brownian
  test particle of mass~$M$ in a thermal bath of temperature~$T$, Eq.~(\ref
  {dpsit2}) leads to the Einstein relation $T/M = D_v/\nu $, where $D_v$~is the
  velocity-space diffusion coefficient.}\BibitemShut {Stop}%
\bibitem [{\citenamefont {Busse}(1967)}]{Busse67}%
  \BibitemOpen
  \bibfield  {author} {\bibinfo {author} {\bibfnamefont {F.~H.}\ \bibnamefont
  {Busse}},\ }\href@noop {} {\bibfield  {journal} {\bibinfo  {journal}
  {J.~Math.\ Phys.}\ }\textbf {\bibinfo {volume} {46}},\ \bibinfo {pages}
  {\range{146}} (\bibinfo {year} {1967})}\BibitemShut {NoStop}%
\bibitem [{\citenamefont {Busse}\ and\ \citenamefont
  {Whitehead}(1971)}]{Busse-Whitehead}%
  \BibitemOpen
  \bibfield  {author} {\bibinfo {author} {\bibfnamefont {F.~H.}\ \bibnamefont
  {Busse}}\ and\ \bibinfo {author} {\bibfnamefont {J.~A.}\ \bibnamefont
  {Whitehead}},\ }\href@noop {} {\bibfield  {journal} {\bibinfo  {journal}
  {J.~Fluid Mech.}\ }\textbf {\bibinfo {volume} {47}},\ \bibinfo {pages}
  {\range{305}} (\bibinfo {year} {1971})}\BibitemShut {NoStop}%
\bibitem [{Note5()}]{Note5}%
  \BibitemOpen
  \bibinfo {note} {Strictly speaking, realizability only requires positive
  semidefiniteness. Sylvester's criterion is then that all of the principal
  minor determinants must be non-negative.}\BibitemShut {Stop}%
\end{thebibliography}%

\end{document}